\documentclass[a4paper,11pt]{article}
\usepackage{aaskaiid}
\usepackage{orcidlink}
\usepackage{aas_macros}
\usepackage[normalem]{ulem}
\usepackage[flushleft]{threeparttable}
\newcommand{\hi}{\mbox{H \textsc{i}}}
\newcommand{\hii}{\mbox{H \textsc{ii}}}

\newcommand{\arcsec}{\mbox{$^{\prime\prime}$}}

\usepackage{color}
\usepackage[colorinlistoftodos,textsize=tiny]{todonotes} 

\title{Spectroscopic surveys with the SKA probing the ionized and molecular Milky Way}
\ShortTitle{The ionized and molecular Milky Way}
\author[1,2]{A.\,Karska \orcidlink{0000-0001-8913-925X}}
\author[3,4]{J.\,R. Dawson\orcidlink{0000-0003-0235-3347}}
\author[5, 2]{A.\,M.\,Jacob \orcidlink{0000-0001-7838-3425}}
\author[6]{T.\,V.\,Wenger\orcidlink{0000-0003-0640-7787}}
\author[7]{J.\,S.\,Urquhart\orcidlink{0000-0002-1605-8050}}
\author[8,2,9]{D.\,Colombo\orcidlink{0000-0001-6498-2945}}
\author[10,11,12]{L.\,D.\,Anderson\orcidlink{0000-0001-8800-1793}}
\author[2,13]{V.\,S. Veena\orcidlink{0000-0002-0801-8550} }
\author[14,15]{M.\,Rugel\orcidlink{0009-0009-0025-9286}}
\author[16,17]{M.\,A.\,Thompson\orcidlink{0000-0001-5392-909X}}
\author[18]{P.\,D.\,Klaassen\orcidlink{0000-0001-9443-0463}}
\author[19]{H.\,Beuther\orcidlink{0000-0002-1700-090X}}
\author[14]{M.\,P.\,Busch\orcidlink{0000-0003-4961-6511}}
\author[20]{A.\,Traficante\orcidlink{0000-0003-1665-6402}}
\author[21]{G.\,Sabatini\orcidlink{0000-0002-6428-9806}}

\ShortName{Karska et al.}
\emailAdd{agata.karska@umk.pl}
\affiliation[1]{Institute of Advanced Studies, Nicolaus Copernicus University in Toruń, Wileńska 4,
87-100 Toruń, Poland}
\affiliation[2]{Max-Planck-Institut für Radioastronomie, Auf dem Hügel 69, 53121, Bonn, Germany}
\affiliation[3]{School of Mathematical and Physical Sciences and Astrophysics and Space Technologies
Research Centre, Macquarie University,
2109, NSW, Australia} \affiliation[4]{Australia Telescope National Facility, CSIRO Space \&
Astronomy, PO Box 76, Epping, NSW 1710, Australia}
\affiliation[5]{ I. Physikalisches Institut, Universit\"at zu K\"oln, Z\"ulpicher Str. 77, D-50937
K\"oln, Germany}
\affiliation[6]{Department of Physics, California State University, Chico, Chico, CA 95929, USA}
\affiliation[7]{Centre for Astrophysics and Planetary Science, University of Kent, Canterbury,
CT2\,7NH, UK}
\affiliation[8]{Argelander-Institut für Astronomie, University of Bonn, Auf dem Hügel 71, 53121
Bonn, Germany}
\affiliation[9]{INAF-Osservatorio Astronomico di Cagliari, Via della Scienza 5, 09047 Selargius
(CA), Italy}
\affiliation[10]{Department of Physics and Astronomy, West Virginia University, Morgantown, WV
26506, USA}
\affiliation[11]{Center for Gravitational Waves and Cosmology, West Virginia University, Chestnut
Ridge Research Building, Morgantown, WV 26505, USA}
\affiliation[12]{Adjunct Astronomer at the Green Bank Observatory, P.O. Box 2, Green Bank, WV 24944,
USA}
\affiliation[13]{Department of Astronomy \& Astrophysics, Tata Institute of Fundamental Research,
Homi Bhabha Road, Mumbai 400005, India}
\affiliation[14]{National Radio Astronomy Observatory, 520 Edgemont Road, Charlottesville, VA 22903,
USA}
\affiliation[15]{Deutsche Zentrum für Astrophysik (DZA), Postplatz 1, 02826 Görlitz, Germany}
\affiliation[16]{School of Physics and Astronomy, University of Leeds, Leeds LS2 9JT, UK}
\affiliation[17]{INAF-Osservatorio Astrofisico di Catania, Via Santa Sofía 78, I-95123 Catania,
Italy}
\affiliation[18]{UK Astronomy Technology Centre, Royal Observatory Edinburgh, Blackford Hill,
Edinburgh EH9 3HJ, UK}
\affiliation[19]{Max Planck Institute for Astronomy, Königstuhl 17, 69117 Heidelberg, Germany}
\affiliation[20]{INAF-IAPS, Via Fosso del Cavaliere, 100, 00133 Rome, Italy}
\affiliation[21]{INAF-Osservatorio Astrofisico di Arcetri, Largo E. Fermi 5, 50125 Firenze, Italy}

\emailAdd{joanne.dawson@mq.edu.au}

\abstract{Radio spectroscopic surveys provide us with a comprehensive picture of the Milky Way
across many physical and chemical regimes. Spectral lines primarily probe the multi-phase gaseous
interstellar medium (ISM) from its ionized to atomic and molecular phases, and constrain both local
and galaxy-scale kinematics and structure through Doppler shifts. By investigating the physical and
chemical properties and distribution of the ISM, the processes driving star formation and galaxy
evolution can be studied in detail. This motivates line surveys of our Galaxy that allow us to use
the range of physical conditions found in the Milky Way as a template for understanding star
formation in extragalactic environments.

In this chapter, we describe the science enabled by the spectroscopy of small molecules and radio
recombination lines of atoms toward a range of Galactic environments with the SKA. We address
questions concerning the processes that dictate the formation of molecular clouds (OH, CH), the
properties of warm, ionized gas and the potential of \hii\ regions in understanding the structure of
the Galaxy (radio recombination lines), and the impact of CO-dark molecular gas across various
density regimes on star formation and galaxy evolution (OH, H$_2$CO). We propose a survey that
includes the inner and outer Galaxy disk, characterized by a broad range of densities, temperatures,
and metallicities. Deep, wide-field observations of small molecules will be uniquely accessible with
SKA, providing key insights on the condition of interstellar medium in galaxies and its impact on
star formation.
}

\begin{document}
\maketitle

\section{Introduction}

The ISM is the lifeblood of galaxies. It provides the reservoir of raw material from which stars are
formed, and to which they return much of their matter at the end of their lives. The cooling and
condensation of warm ISM into cold molecular clouds directly determines conditions for star
formation, a process that is driven both by galaxy-scale effects and by small-scale physics and
chemistry. 
As young stars evolve, they heat and ionize the ISM, and enrich it with the dust and heavy elements responsible for cooling and shielding. The exchange
of matter and energy between stars and the ISM thus forms a complex and interconnected system of
self-regulating feedback that underpins the evolution of galaxies.

Our own Milky Way is an ideal laboratory for studying these connections and their dependence on
environment in unprecedented detail, and provides a template for understanding the star formation
and ISM lifecycle in extragalactic environments. Spectroscopic surveys are a key part of this picture -- essential to understanding our Galaxy's
gas-phase constituents, linking the properties of the ISM to the processes of star formation, and
connecting physical processes to  environment. They provide kinematic information for the ionized,
atomic and molecular structures of the Milky Way, revealing the physical and chemical processes
underway as well as the distances to astronomical objects, all on spatial scales generally
unobtainable in external galaxies. Combined with high-resolution continuum surveys, they are also
uniquely suited to mapping the three-dimensional structure of the Galaxy in unprecedented detail. In
this chapter, we focus on surveys of the ionized and molecular constituents of the Milky Way, noting
that the atomic component is addressed in detail in \cite{Miville-Deschenes01.2026.SKA}.

A key transition point in the stars--ISM lifecycle is the formation of molecules in cold,
filamentary atomic clouds \citep{krumholz2009}. Fundamentally, the ISM becomes molecular once
density, temperature, and column density thresholds are met, which can occur due to both local and
galaxy-scale processes. However, the details of how molecular cloud formation occurs, both on small
scales and in the galactic context, remain uncertain. In particular, the connection between the
physical conditions in the atomic gas and the propensity and mechanisms of molecule formation is not
well constrained observationally \citep[e.g.][]{hafner2023,park2023}; nor are the details of
molecular cloud assembly on sub-parsec scales. Additionally, there is growing evidence that H$_2$
exists at relatively low abundance throughout much of the cold atomic medium and is undetectable in
the CO lines most commonly used to trace the molecular ISM
\citep{liszt2000,busch2021,rybarczyk2022}. This apparently ultra-diffuse molecular component is
still difficult to detect. However, it is likely the tail end of a much larger reservoir of diffuse,
so-called `CO-dark' molecular gas \citep[e.g.][]{grenier2005}, which may make up a considerable
fraction of the H$_2$ budget, particularly in the outer Galaxy and in cloud envelopes
\citep[e.g.][]{pineda2013,remy2019}. While CO-dark H$_2$ is unlikely to participate directly in star
formation, it is critical as a marker of the earliest stages of chemical complexity, and may be
important for the total molecular mass budget, particularly in the metal-poor outer Galaxy.

\begin{figure}[ht!]
    \centering
    \includegraphics[width=1\linewidth]{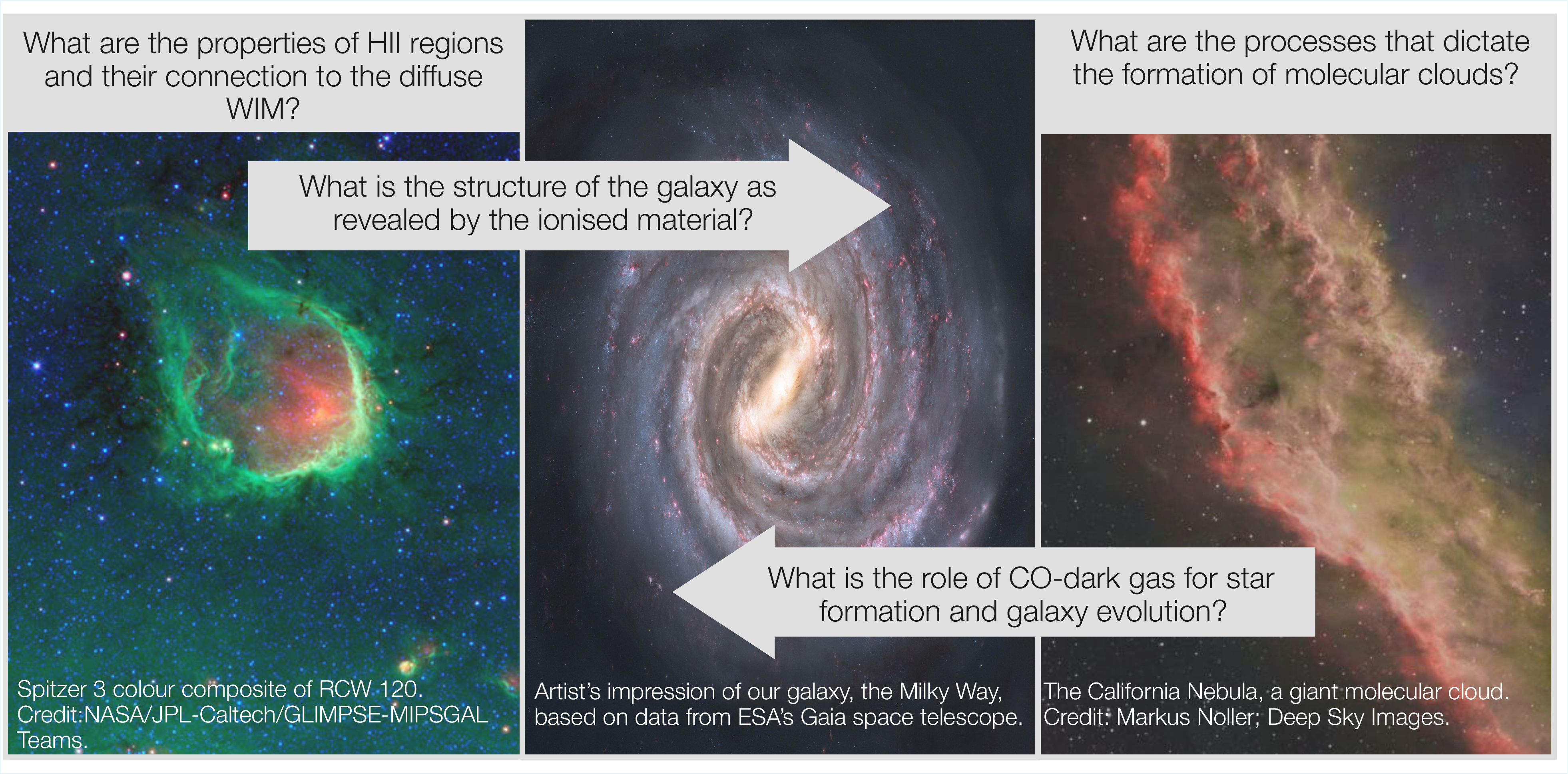}
    \caption{The primary scientific goals of this work are illustrated here, highlighting the
multi-scale and multi-physics impact of spectral line surveys.}
    \label{fig:overview}
\end{figure}

Within the Solar circle, roughly half of the interstellar gas mass resides in the molecular phase
\citep{sco87,Bro88,heyer_dame2015}. However, the properties of the molecular gas and its ability to
form stars change as a function of Galactocentric radius ($R_\mathrm{Gal}$), from the Galactic
center to the inner and outer Galaxy ($R_\mathrm{Gal}>8.2$ kpc; \citealt{gravity19}). In particular,
the gradual decrease in metallicity \citep[e.g., ][]{men22} results in reduced abundances of dust
and molecules heavier than H$_2$ in the outer parts of the Milky Way
\citep{Ngan2023,Patra2025,Gigli2025}. The average flux of cosmic rays and the ultraviolet radiation
field are also reduced \citep{Bl87,padovani2020, Sabatini23}, but both penetrate deeper into clouds
due to the reduced shielding, where they can heat and ionize gas \citep{ossenkopf2025}. The
decreased efficiency of gas and dust cooling affects the condensation and fragmentation of molecular
clouds \citep{glover12}, and combined with the high fraction of atomic-to-molecular gas, is expected
to lower the star formation rate per surface area \citep{ke12}.

So far, spectroscopic surveys of the Galactic plane have mainly used carbon monoxide (CO) as the
tracer of the bulk of the molecular gas mass. In the inner Galaxy, CO surveys have delivered a
comprehensive census of molecular clouds \citep{roman-duval10,barnes15,duarte-cabral2021}, mapped
their association with spiral arms \citep{roman-duval10,colombo22}, and studied the connection
between clump properties and star formation \citep{eden2013,urquhart2021}. Similar surveys in the
outer Galaxy, however, have shown a weaker association of star formation with spiral arms
\citep{koenig2021,urquhart24,urquhart25}, which may be related to the presence of supershells or
other dynamical processes within the disk, such as flaring and scalloping, as revealed by \hi\
observations \citep{kalberla09,soler22}. Since CO can be efficiently photodissociated by UV
radiation from young stars, 
alternative tracers are necessary to reveal the hidden, CO-dark molecular gas in the ISM and its
relation to star formation. 

Once stars have formed, they have an immediate effect on their surroundings, in particular via the
intense ionizing radiation emitted from the most massive stars. The $\sim\!10^4$\,K ionized ISM
consists of \hii\ regions surrounding high mass stars and a pervasive, diffuse Warm Ionized Medium
(WIM) that fills the volume of the disk and extends far from the plane \citep{lyne85}. As signposts
of active star-formation, \hii\ regions have historically been used as tracers of the Milky Way's
spiral structure \citep{georgelin_georgelin1976}. While the dominant source of ionization for the
WIM is believed to also be high mass stars \citep{haffner09}, confirmation requires connecting the
plasma properties and distribution in the dense zones surrounding \hii\ regions with those in the
diffuse WIM. 

The diffuse WIM can be detected spectroscopically using optical lines such as H$\alpha$, but
extinction severely limits the utility of such measurements in the Galactic plane. 
Radio recombination lines (RRLs), primarily of H but also He and C, are detected throughout the
entire volume of the Galactic disk \citep{anderson21,2024A&A...689A..81K} and provide velocities to
\hii\ regions that, when combined with models of Galactic rotation, can result in kinematic distance
measurements. They are very effective tracers of the $\sim\!10^4$\,K ionized ISM, 
providing constraints on the star-formation and stellar feedback processes via measurements of
electron temperatures, the level of turbulence and (indirectly) metallicity, and the association
between ionized gas and natal molecular material \citep[e.g.,][]{2017MNRAS.465.4219V}. 

Large-scale line surveys of the Milky Way with the SKA provide the opportunity to study the physics
of the ISM and its links to star formation using several complementary species: hydroxyl (OH),
methylidyne (CH), formaldehyde (H$_2$CO), and radio recombination lines of hydrogen, helium, and
carbon (H, He, C). In this chapter, we propose an SKA survey that allows us to link the evolution of the ISM with star
formation and galaxy evolution. The key science themes include: (i) understanding how cold atomic
gas conditions impact molecular gas formation as a function of $R_\mathrm{Gal}$, including the role
of CO-dark gas, (ii) characterizing the warm, ionized gas surrounding \hii\ regions and its
connection to the diffuse WIM, (iii) constraining the structure of the Galaxy using kinematics from
radio recombination lines, and (iv) establishing the role of CO-dark gas across various density
regimes, constrained with H$_2$CO, for star formation and galaxy evolution (see also
Fig.~\ref{fig:overview}).

In Section \ref{sec:cnm_to_mgas}, we describe the science motivation for the line survey of the
Milky Way. Section \ref{sec:tracers} provides a description of each of the tracers and their utility
for the proposed science. In Section \ref{sec:survey}, we discuss the potential scope of the survey
in terms of the coverage within the Milky Way and its feasibility. In Section \ref{sec:summary}, we
provide a summary and outlook.

\section{Linking the evolution of the ISM with star formation and galaxy evolution}
\label{sec:cnm_to_mgas}

The ISM exhibits a complex thermal, ionization, and density structure with multiple phases in
approximate pressure equilibrium \citep{mckee_ostriker1977}. These include: the hot ionized medium
(HIM; $T \sim 10^6$~K; \citealt{cox_smith_1974}), the warm ionized medium (WIM; $T \sim 10^4$~K;
\citealt{reynolds1983}), the warm neutral medium (WNM; $T \sim 10^4$~K;
\citealt{kulkarni_heiles1987,dickey_lockman1990}), the cold neutral medium (CNM; $T \sim 100$~K),
and molecular clouds ($T \sim 10$~K).
The flow of material between these phases reflects the balance of heating and cooling, and the
energy input from stars. Transitions in and out of the densest, coldest molecular gas phase directly
determine star formation rates. 

\begin{figure}[ht!]
     \centering
     \includegraphics[width=1\linewidth]{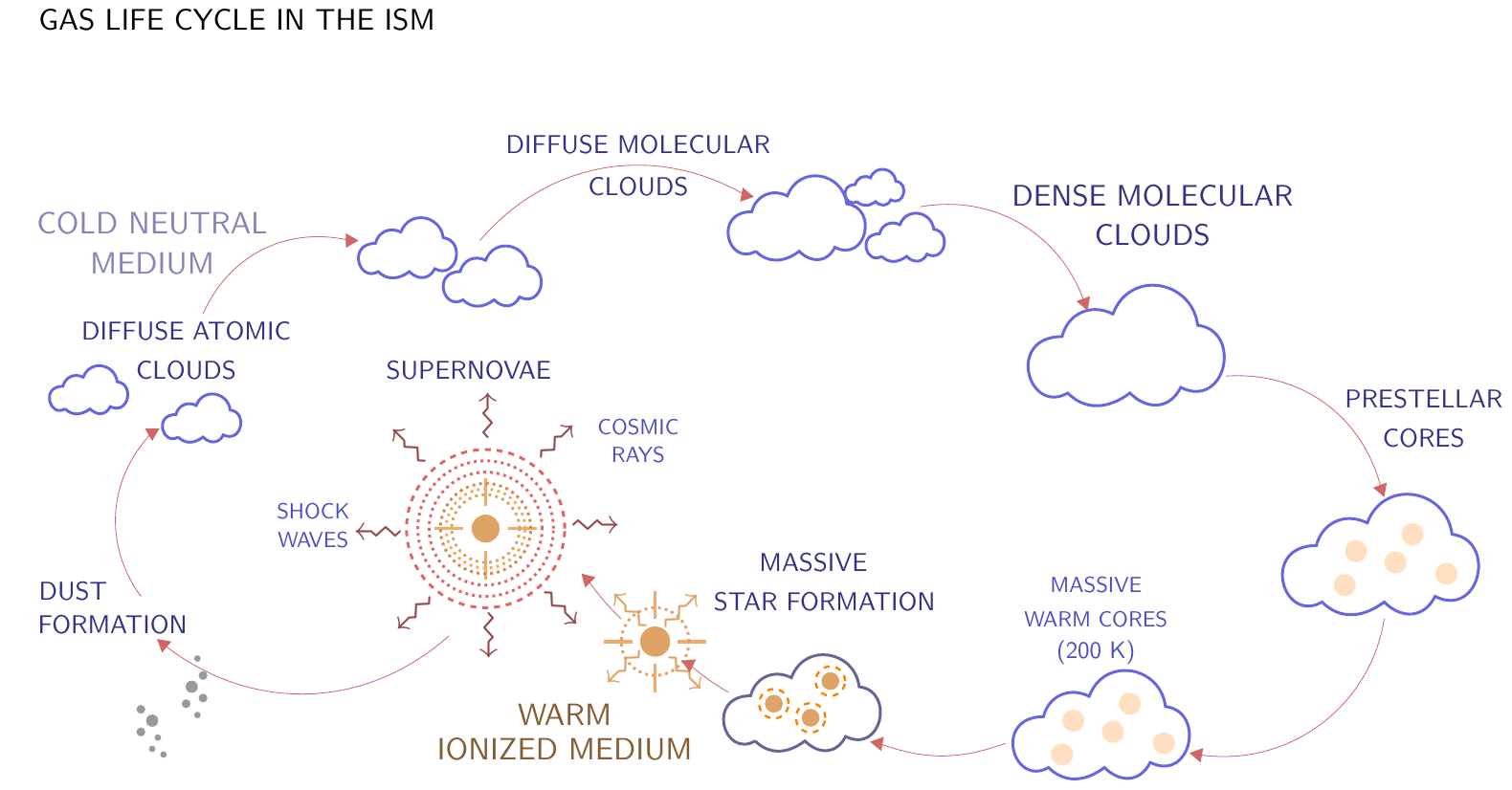}
     \caption{Schematic illustrating the lifecycle of baryonic material in the ISM, tracing its
evolution from diffuse atomic gas to molecular clouds, the collapse into dense star-forming regions,
the birth and evolution of massive stars, and their eventual return of enriched material back to the
ISM.}
     \label{fig:enter-label}
 \end{figure}
\subsection{From cold atomic to molecular gas}
\label{sec:2.1}

In the interests of clarity, we will begin by defining terminology. The term `cold neutral medium'
(CNM) is used variously in the literature to refer to anything from purely atomic cold gas, to all
ISM below $\sim$100\,K, including both diffuse-molecule dominated regions and dense star-forming
molecular clouds. Here we will use CNM to refer to cold gas that is \textit{primarily} atomic, but
that may contain molecules at low abundance. This reflects historical usage of the term to refer to
one of the two thermally stable phases of the atomic ISM \citep[in opposition to
WNM,][]{wolfire1995}, and a rich history of its study via the \hi\ 21\,cm line
\citep[e.g.][]{heiles2003}. 
However, we acknowledge that even primarily-atomic ISM may contain a non-negligible molecular
fraction \citep[see e.g.][]{busch2021,rybarczyk2022}, noting that the process of molecule formation
can begin at relatively low column density \citep[e.g.][]{wolfire03} and that molecules may be
stirred into otherwise atomic gas via turbulent mixing \citep[e.g.][]{valdivia2016}.
In our definition, CNM does \textit{not} include H$_2$-dominated regions -- these we refer to as
`molecular gas' or `molecular clouds'. But we note that CNM structures may host molecular gas within
them.

With terminology defined, we begin by recognizing that molecular clouds form when the cold atomic
medium accumulates to sufficiently high volume and column densities to (self-)shield against
photodissociation, enabling the formation and survival of H$_2$
\citep{draine1996,Krumholz2008,krumholz2009,glover2011,sternberg14}. This process may be driven by
various mechanisms, including turbulent flows, cloud mergers, and gravitational instabilities, 
and is governed by environmental factors such as pressure, metallicity, and dust shielding, 
as well as local physical conditions
\citep{elmegreen89,blitz_rosolowsky2006,leroy2008,krumkholtz10,sternberg14}. Since molecular clouds
require CNM as a precursor to their formation, examining the relationship between the physical state
of the CNM and the properties of 
any associated molecular gas is critical to understanding the conditions for star formation (Fig.
\ref{fig:enter-label}). This must be done across a range of column density regimes, from low $A_V$,
partially molecular gas, to dense and well-shielded molecule-dominated regimes, across a range of
Galactic environments, and with sufficient number statistics to deduce robust correlations from
inherently scattered data.

While the galaxy-scale processes responsible for molecular cloud formation are relatively well
understood \citep[e.g.][]{krumholz2009}, and the global column density thresholds governing the
H{\sc i}-to-H$_2$ phase transition are well constrained \citep[e.g.,][]{bellomi2020}, the manner in
which the transition occurs on small (sub-pc to pc) scales remains less clear. Yet these are
precisely the scales on which many of the key physical processes shaping the interstellar medium
operate. For instance, the rich $\sim$pc-scale filamentary structure observed in the CNM
\citep[e.g.,][]{mcclure2006,clark2019} is thought to arise from thermal instabilities in the
magnetized, turbulent ISM, which naturally organize the CNM into filaments and clumps on
characteristic scales of $\lesssim$0.1–10~pc \citep{audit2005,inoue2016}. However, while structure
formation within the CNM sets the starting conditions from which molecular clouds are born, we lack
a good description of how CNM structure couples to the star formation process; for example, it is
unclear if atomic filaments are the direct precursors of self-gravitating molecular filaments that
are widely observed in star forming regions \citep{arzoumanian11,andre2014,hacar2023}. Similarly,
while the physical conditions required for molecule formation must arise from the physics of the
atomic medium, our ability to link the thermal, structural and chemical properties of 
associated atomic and molecular gas is still limited -- both by sample size and complexity
\citep[e.g.][]{syed2020,hafner2023,park2023,soler23}.

\begin{figure}[ht!]
    \centering
    \includegraphics[width=0.8\linewidth]{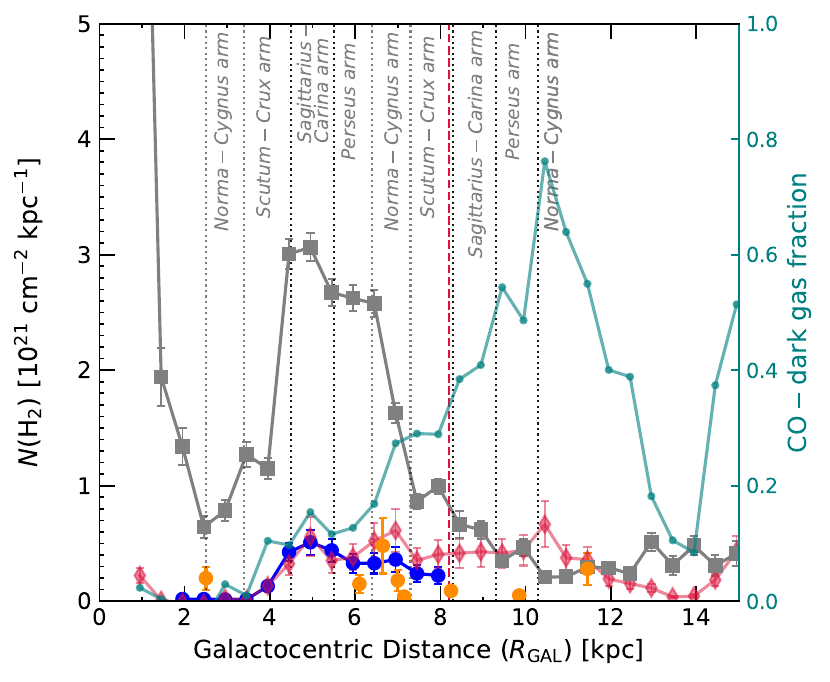}
    \caption{Radial distribution of CO-dark molecular gas column density in the plane of the Milky
Way as traced by varied tracers including [\mbox{C \textsc{ii}}] (pink diamonds), CH (blue circles),
and OH (orange circles) alongside the total H$_2$ column density as traced by CO (gray squares). The
spiral arm crossings, typically encountered along the line of sight toward a background continuum
source in the fourth Galactic quadrant, are indicated assuming a spiral-arm width of 0.85~kpc. The
vertical dashed red line marks the Sun’s galactocentric distance. The secondary y-axis (in teal)
displays the CO-dark gas fraction as traced by [\mbox{C \textsc{ii}}]. Adapted from
\citet{Jacob2023}. Data sources: CO, [\mbox{C \textsc{ii}}], CO-dark gas – \citet{pineda2013}; CH –
\citet{Jacob2019}; OH - \citet{Jacob2019, busch2021}.
}
    \label{fig:Compare_CO-dark-gas_tracers}
\end{figure} 

One key place we lack understanding is  the role of diffuse, CO-dark H$_2$ in molecular cloud
formation. Often referred to as `dark molecular gas' (DMG), CO-dark H$_2$ exists in warmer
low-extinction environments ($A_V \lesssim 1$), where limited (self-)shielding prevents the survival
of a substantial amount of CO 
\citep{grenier2005,wolfire2010, shull2021}, but the transition from \hi\ to H$_2$ is well underway.
The fraction of CO-dark H$_2$ to total H$_2$ in the ISM ranges from $\sim$20\% at $R_\mathrm{GC}$ of
4 kpc to $\sim$80\% at 10 kpc (Figure \ref{fig:Compare_CO-dark-gas_tracers}), and significant areas
of the sky associated with OH emission are CO-dark \citep[e.g.][]{pineda2013,busch19, busch2021}.
Regions dominated by DMG may mark the earliest stages of molecular cloud formation, but DMG also
exists in the envelopes of CO-bright clouds \citep[e.g.][]{traficante14,remy2019,liszt2019},
and may even be well-mixed with \hi\ \citep{liszt2000,busch2021,rybarczyk2022}. How such mixing
could be achieved is an open question. Numerical models suggest that transient turbulent
overdensities allow for rapid formation of H$_2$, which may then be mixed into the more diffuse
surroundings \citep[e.g.][]{glover2007, valdivia2016}. However, observational results so far are
unable to distinguish between this scenario and the presence of tiny, unresolved dense cloudlets
\citep[e.g.][]{pringle2001}. In any case, the distribution of DMG on these small scales has direct
implications for models of molecular cloud assembly. 

Figure~\ref{fig:Compare_CO-dark-gas_tracers} illustrates the radial distribution of the CO-dark gas 
across the Galactic plane as traced by [\mbox{C \textsc{ii}}], CH, and OH, alongside the total
molecular gas content traced by CO. These comparisons clearly demonstrate that the CO-dark fraction
(relative to the total molecular content) is significant along many sightlines, with especially
large fractions in the outer Galaxy. The [\mbox{C \textsc{ii}}] dataset is extensive -- observed
under the Herschel/HIFI key guaranteed program, GOT C+ \citep{pineda2013}, covering $\sim$500
Galactic sightlines. However, since [\mbox{C \textsc{ii}}] exists in all phases of the ISM, there
are challenges in isolating the CO-dark component from [\mbox{C \textsc{ii}}] alone. The molecular
species CH and OH have advantages in this regard, but measurements of both remain limited. A large
fraction of the available CH and OH data come from absorption line studies of their FIR rotational
lines observed with SOFIA \citep{Jacob2019}, and are restricted to only a handful of sightlines. OH
in particular presents additional challenges, as its FIR lines are often optically thick. The radio
OH lines at 18 cm have been extensively observed with single dish facilities
\citep[e.g.][]{dawson2014,dawson2022,xu2016,li2018,busch19,busch2021} and interferometers
\citep[e.g.][]{rugel2018,rugel2025,hafner2023}, and are able to trace CO-dark H$_{2}$. However,
single-dish studies within the Galactic Plane present challenges due to in-beam and line-of-sight
confusion \citep{dawson2022}, and existing interferometric observations have generally not been
sensitive enough to detect CO-dark gas. Sensitive OH and CH observations with the SKA, described
below, will therefore be crucial to characterizing the nature and distribution of this diffuse
molecular component.

\subsection{The Warm Ionized Medium}

The warm ionized medium (WIM) is the $\sim\,10^4$\,K ionized gas that pervades the volume of the
Galaxy. Typically residing far from the Galactic Plane, a diffuse WIM component with densities as low as
$n_e\simeq\!10^{-2}\,{\rm cm}^{-3}$ \citep{lyne85} is detected using optical lines such as H$\alpha$
\citep{haffner03} or via pulsar dispersion measures. The densest portions of the WIM -- the ``dense
warm ionized medium'' \citep[D-WIM;][]{langer21} -- are found just outside of \hii\ regions, with
electron densities 10--30 cm$^{-3}$, similar to those of \hii\ regions
\citep{roshi97,heiles01,goldsmith15}, and containing up to a third of the WIM mass. 

One of the major questions surrounding the WIM is the source of ionization. 
Based on energy considerations, the ionization of the WIM is most likely maintained by OB stars.
\citet{haffner09} found that all of the energy from Galactic supernovae is required to maintain the
WIM ionization, but only 15\% of the energy from OB stars is required. The OB star hypothesis must
be able to explain how OB star photons escape from their dense natal environments to reach large
distances above the plane, `leaking' through \hii\ region photodissociation regions (PDRs)
\citep[e.g.,][]{lockman76, anantharamaiah86}. 
The spectrum of the WIM differs from that of \hii\ regions, however: it has characteristically lower
[O{\sc iii}] / H$\alpha$ and He{\sc i} / H$\alpha$ ratios as well as higher [N{\sc ii}] / H$\alpha$
and [S{\sc ii}] / H$\alpha$ ratios. Thus, if OB stars provide the requisite photons, the hardness of
the radiation field must be modified as the photons leak from \hii\ regions. To determine the source
of ionization and how it is being softened, we need to trace the WIM from \hii\ regions into more
diffuse environments; with these observations we could show that the plasma properties can be
explained as coming from individual \hii\ regions \citep[cf.][]{luisi16}.

Our present understanding of the WIM has been hampered by the lack of a facility that can connect
these two regimes. Optical observations of the WIM are very sensitive, but suffer from extinction at
locations where the D-WIM is found. Radio observations have lacked the necessary sensitivity, and so
can detect the D-WIM, but not the more diffuse WIM. We need sensitive radio observations to
investigate whether the WIM components have a common origin, determine their distribution, and to
determine their plasma properties.

A connection between the D-WIM and \hii\ regions was investigated by \citet{anderson15c}, who found
that the locations of large star-forming complexes in the inner Galaxy are correlated with the
detection of multiple ionized gas components, which they interpreted as indicating the presence of
the (dense) WIM preferentially toward massive star forming regions. \citet{luisi20} analyzed maps of
RRL data and found that they could reproduce the extended ionized gas emission around the W43 star
forming complex by assuming that all \hii\ regions in the vicinity leak a percentage of their
photons into the ISM. Because the kinetic temperature of the D-WIM is similar to that of the more
diffuse WIM, the D-WIM is at a higher pressure.

The distribution of the WIM is also unclear, although it is well understood that the bulk of the
dense gas is associated with high-mass star formation regions. Within the plane of the Milky Way,
there appears to be a low-temperature (kinetic temperatures $<\!4,\!000$\,K) component that is not
associated with star formation. In a deep survey, taking $\sim\!20$\,hours on source and
$\sim\!200$\,hours effective after stacking 15 C-band RRLs, \citet{bania24} detected inner Galaxy
diffuse ionized gas whose narrow line width implies very low electron temperatures. The emission may
arise from cool diffuse plasma entrained in the Galactic bar, which would explain why it has no
obvious sources of ionization, but does not provide an explanation for its cool temperature. To
determine the extent and reasons for the low temperature of this component, very sensitive radio
observations are required that are capable of tracing the ionized gas properties over a wide range
of electron densities.

\begin{figure}
    \centering
    \includegraphics[width=\textwidth]{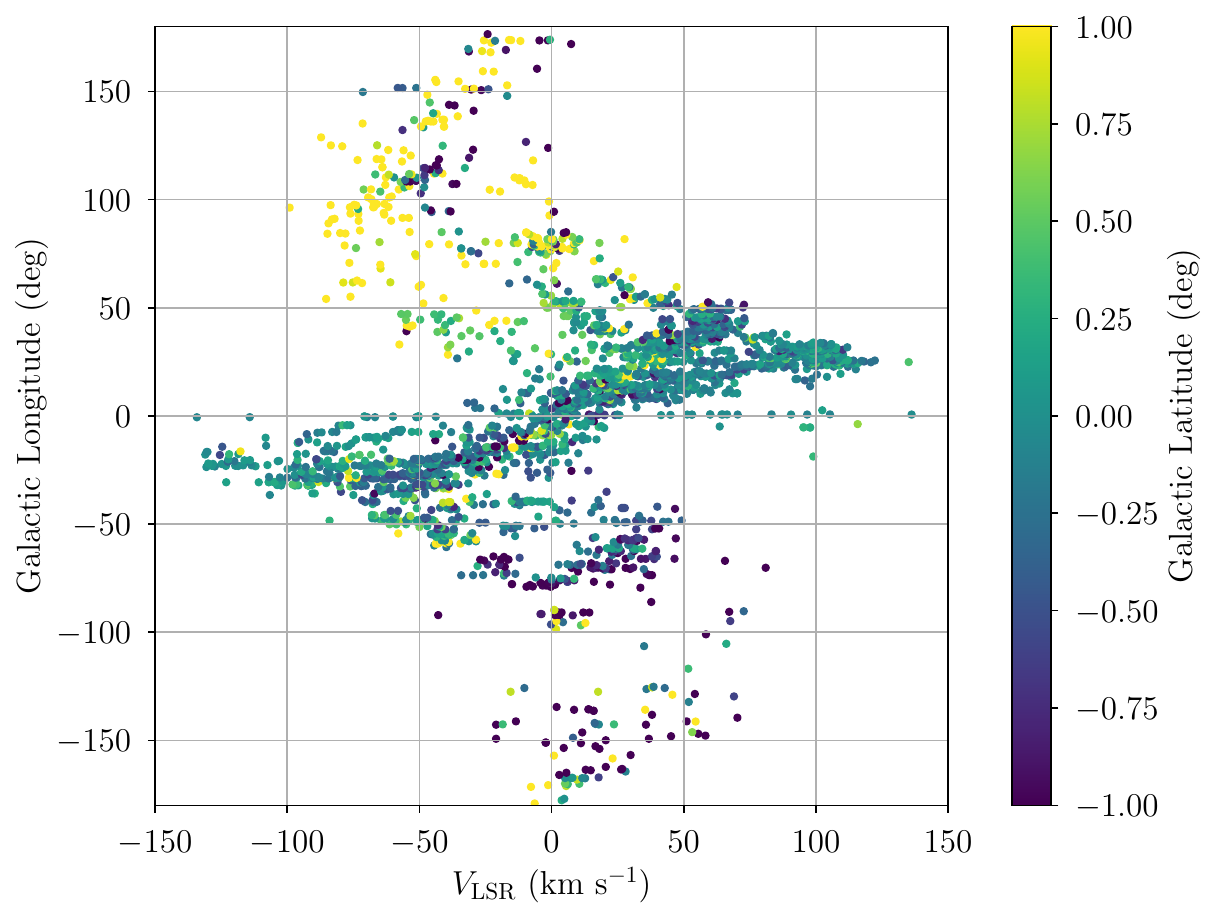}
    \caption{\label{fig:hii_lv} Galactic longitude, latitude, and LSR velocity ($V_{\rm LSR}$)
distribution of 2,582 spectroscopic ionized gas measurements associated with 2,351 Galactic \hii\
regions in the \textit{WISE} Catalog of Galactic \hii\ Regions \citep{anderson14}. An additional 21
measurements with $|V_{\rm LSR}| > 150$ km s$^{-1}$ toward 20 nebulae (one at a Galactic longitude
near $\ell = 7.70^\circ$, the rest near $\ell = 358.75^\circ$) are excluded for clarity. Along
sight-lines with multiple ionized gas velocity components, one or more of those components likely
originates in diffuse ionized gas rather than a discrete \hii\ region \citep{anderson15b}.}
\end{figure}

\subsection{\hii\ regions: deciphering Galactic structure}

\hii\ regions are formed around high-mass OB stars and are the classic tracer of spiral structure in
galaxies \citep{rosse1850}. In the Milky Way, the positions and kinematics of \hii\ regions reveal
structures that are interpreted as spiral arms \citep[e.g.,][]{georgelin_georgelin1976,hou_han2014}.
For most Galactic \hii\ regions, the distances to the nebulae are estimated kinematically by
assuming a model of Galactic rotation \citep[e.g.,][]{anderson12c}. These kinematic distances are
subject to large, systematic uncertainties \citep{wenger18,peek2022}. The three-dimensional position
and kinematic information provided by maser parallax observations constrain both the morphological
structure as well as the kinematic structure of the Galactic disk \citep{reid14,reid2019}. However,
such measurements are only available for a few hundred nebulae, and are almost entirely confined to
sources visible from the Northern hemisphere, due to the lack of Southern hemisphere facilities.

In the absence of parallax measurements or other independent distance estimates, the
position-position-velocity (ppv) distribution is the fundamental observable dataset for Galactic
structure tracers like \hii\ regions and \hi\ emission. For example, Figure \ref{fig:hii_lv} shows
the Galactic longitude-velocity distribution of every nebula in the \textit{WISE} Catalog of
Galactic \hii\ Regions \citep{anderson14}. This distribution encodes both the underlying
three-dimensional structure of the tracer as well as the velocity structure along the line-of-sight.
Traditionally, the kinematic distances to structure tracers are inferred under the assumption of an
axisymmetric rotation curve \citep{wenger18}.  Deviations from a simple Galactic rotation model, as
seen in the Perseus arm, are known to systematically bias kinematic distance estimates and create
``features'' in the ppv distribution \citep{peek2022}. Such features correspond to velocity caustics
rather than true enhancements in the tracer density. With (1) a sufficient sample of Galactic
structure tracers like \hii\ regions, (2) a sophisticated model of Galactic morphological and
kinematic structure, and (3) probabilistic inference tools like Markov chain Monte Carlo (MCMC)
techniques, the degeneracies between three-dimensional density enhancements and velocity caustics in
the ppv distribution could be broken {\it probabilistically}. Sufficiently sampled, the ppv
distribution {\it constrains} models of Galactic structure.

In the inner Galaxy, an axisymmetric Galactic rotation curve causes an ambiguity in the inferred
kinematic distance. There are two distinct distances at which matter is predicted to have the same
radial velocity. This problem is known as the kinematic distance ambiguity (KDA). Since \hii\
regions emit broadband thermal radio continuum, the presence of absorbing material in the foreground
of an \hii\ region can resolve the KDA \citep{anderson09a,brown2014}. Given the prevalence of cool
\hi\ in the Milky Way disk, \hi\ emission/absorption observations are commonly used to resolve the
KDA \citep{kuchar94,kolpak03,urquhart12}. Unless the KDA is resolved, the kinematic distances of
\hii\ regions are worse than unreliable: they are degenerate.

The continuum survey with SKAO in Band 5b covering the 10-15 GHz frequency range with a sensitivity
of approximately 20\,$\mu$Jy and a resolution of $\simeq0.07$\arcsec is designed to identify
hundreds of thousands of compact radio sources \citep{Traficante01.2026.SKA}. In particular, these
catalogs will be sensitive enough to detect all \hii\ regions across the Galactic disc from the
youngest and smallest ones, the so-called hyper-compact \hii\ regions, up to the more evolved and
extended classical \hii\ regions, providing a complete census of their population as well as their
evolution. In addition, the statistical lifetimes of their early stages will be robustly determined.

However, \hii\ regions are expected to account for only about 1\% of the compact radio sources
detected in these continuum surveys; thus, classifying them requires additional information.  
RRLs observed with SKA will play a crucial role in confirming \hii\ region candidates identified
from positional correlations with infrared and submillimeter catalogs (e.g., WISE
\citealt{anderson14}; and Hi-GAL; \citealt{elia2017,elia2021}) and in their subsequent
characterization. The detection of an RRL confirms that a source is Galactic, allows the distance to
the \hii\ region to be determined, and thus establishes its location in the Galaxy. The
line-to-continuum ratio can be used to derive the electron temperature, while the line width
provides information on the ionized gas kinematics and pressure broadening effects. The velocity of
the line emission also makes it possible to identify the natal molecular material and to study the
interaction between the molecular and ionized gas.

\subsection{The Milky Way's star formation in the context of galaxy evolution}
\label{sec:starformation}

The formation of stars occurs in molecular clouds that constitute the coldest and densest part of
the ISM in galaxies, with kinetic temperatures $T_\mathrm{kin}\sim10$ K and H$_2$ number densities
$n_\mathrm{H_2}>30$ cm$^{-3}$ \citep{ke12}. Gravitational collapse of, and turbulent mixing in,
molecular clouds leads to fragmentation, and the formation of smaller substructures: clumps -- the
birthplaces of stellar clusters ($n_\mathrm{  H_2}\sim10^4-10^5$ cm$^{-3}$; sizes $\sim0.7$ pc,
\citealt{Traficante15,urquhart18}), and dense cores, forming individual stars or close multiple
systems of stars within those clumps ($n_\mathrm{ H_2}>10^5$ cm$^{-3}$; sizes $<0.1$ pc,
\citealt{pineda23}). The rate of star formation is closely linked with the available reservoir of
molecular gas, and this relationship seems to hold all the way from the size scales of dense clumps
in the Milky Way \citep{Wu05,Wu10} to those of starburst galaxies \citep{gao04}.

Nevertheless, recent cloud-scale nearby galaxy observations (from e.g., the `Physics at High Angular
resolution in Nearby GalaxieS', PHANGS survey) have shown that star formation is strongly regulated
by the local environment, rapid cloud evolution, and stellar feedback \citep{schinnerer_leroy2024}. 
Molecular clouds  appear short-lived, with lifetimes of roughly 10 to 30 Myr, inferred from spatial
offsets between CO, H$\alpha$, and UV emission, implying rapid cycling between cloud assembly, star
formation, and dispersal \citep[e.g.][]{kruijssen2019, chevance20}. Stellar feedback from young
clusters plays a central role in dispersing clouds and regulating the star formation process,
preventing runaway collapse and maintaining low integrated efficiencies \citep[e.g.][]{kim2022}. As
a result, only a few percent of molecular gas is converted into stars per free-fall time, allowing
galaxies to sustain star formation over Gyr timescales rather than exhausting their reservoirs
rapidly \citep[e.g.][]{sun2020,leroy2021}.
 
These low SFEs have been observed to be mostly connected to the effects of large-scale dynamics,
such as spiral arm streaming motions \citep[e.g.][]{meidt2013}, bar inflows
\citep[e.g.][]{egusa2018,maeda2020,hogarth2024}, shear in galaxy centers
\citep[e.g.][]{leroy2013,utomo2017,hatchfield2020,henshaw2023}, and gravitational potentials from
spheroids \citep[e.g.][]{colombo2018} that increase the molecular cloud stability threshold and
reduce gas fragmentation. Energetic processes, including active galactic nuclei and feedback from
massive stars, can also contribute by heating the gas to temperatures unsuitable for the formation
of stars. Indeed, the interplay between molecular gas availability and the ability to turn this gas
efficiently into new stars is one of the main factors that establish the evolutionary stage of a
galaxy \citep[e.g.][]{saintonge2017,colombo2020,villanueva2023,pan2024,colombo2025b}. However, to
establish which effect dominates requires a precise knowledge of the bulk of molecular gas available
in a given star forming region in the Milky Way. 

\begin{figure}
     \centering
     \includegraphics[width=1\linewidth]{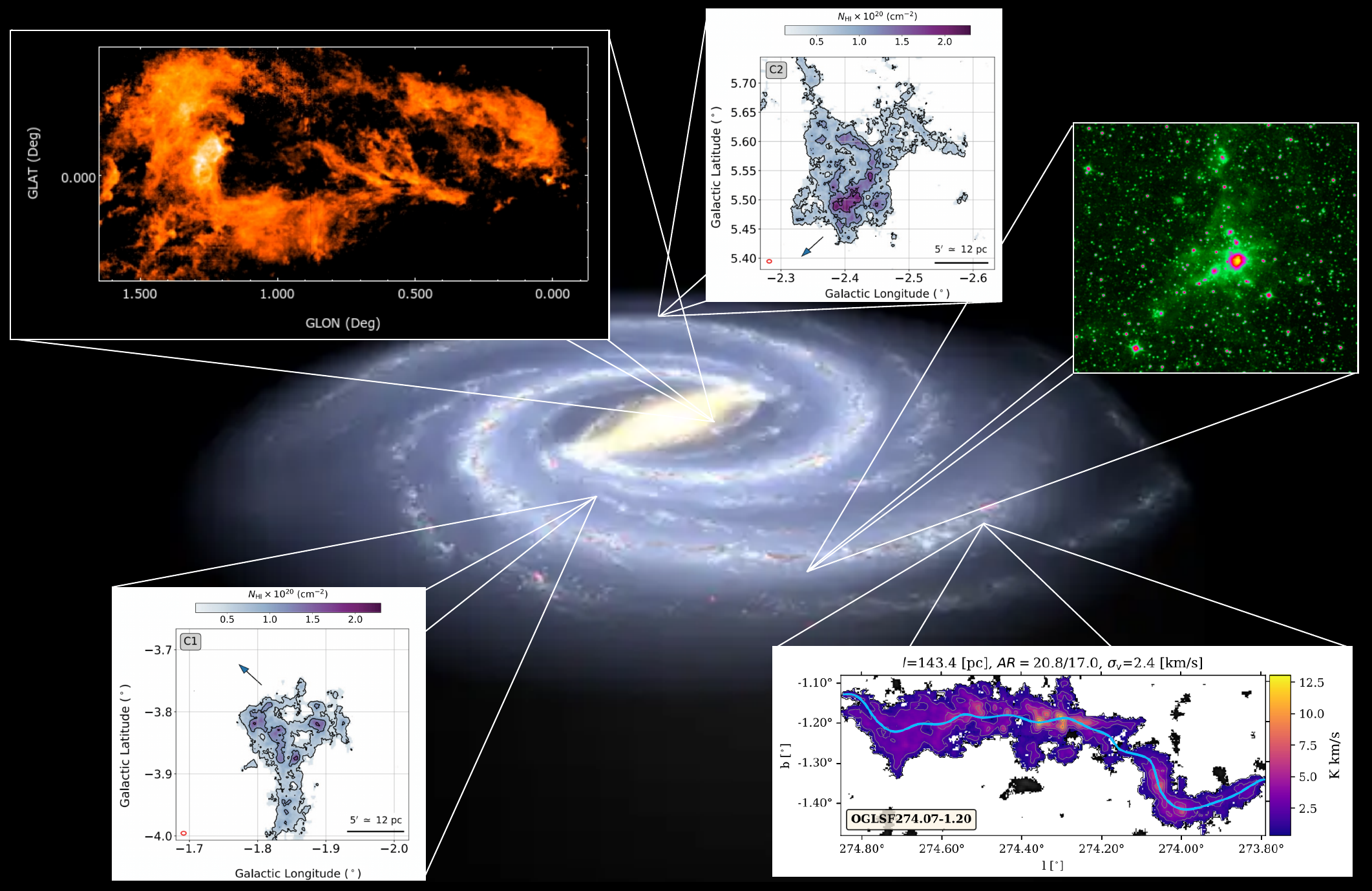}
     \caption{Galactic environments spanning the metallicity and dynamical diversity of the Milky
Way, so far observed only in the CO-bright gas, forming the target sample for the proposed SKA line
survey. Clockwise from top-left: high-velocity gas stream tracing bar-driven inflow in the Central
Molecular Zone \citep{2024A&A...689A.121V}; an HVC toward the northern lobe of the nuclear outflow
\citep[][]{2023MNRAS.524.1258N}; the G195.722-0.120 high-mass star-forming complex in the far outer
Galaxy (Urquhart et al. 2026, in prep.); the Falkor large-scale molecular filament in the outer
Galaxy (OGLSF274.07-1.20, \citealt{colombo21}, Rodrigues da Costa et al. in prep.); and an HVC
toward the southern lobe of the nuclear outflow \citep[][]{2023MNRAS.524.1258N}.}
     
     \label{fig:targets}
 \end{figure}

Placing our understanding of Galactic star formation into its extragalactic context, we find that
the Milky Way might not be the most efficient star forming engine, but instead a galaxy 
that is slowly reducing its star formation activity, on the way to becoming a retired object
\citep[e.g.][]{mutch2011,licquia2015}. This could stem from the dichotomy of observed inner and
outer Galaxy behaviors: while the inner Galaxy 
appears to be actively forming stars, the outer Galaxy is generally quiescent. In fact, the Milky
Way hosts a number of environments spanning a range of metallicity and dynamical properties such as
the nuclear outflow, bar-driven inflow, and far outer Galaxy filaments and clumps (see
Fig.~\ref{fig:targets}).
Therefore, by looking at the Milky Way we have the opportunity to understand not only how stars
form, but also how stars stop forming, peering into one of the most important galaxy evolution
processes at the highest level of detail possible. 

We begin with the generally quiescent outer Galaxy, where a significant amount of molecular gas
might be CO-dark. The sub-solar metallicity allows for less shielding, and greater penetration of UV
photons. The consequences of this are warmer dust grains \citep{vloon2010} and more efficient
photodissociation of molecules such as CO \citep{wolfire03}.
This effect was demonstrated by far-IR observations of the \hii\ region 30 Doradus in the Large
Magellanic Cloud (LMC), where 75\% of the molecular gas mass is suggested to be undetected with CO
\citep{chevance20}. The outer Galaxy has a similar metallicity to the LMC, but an at least
10--100$\times$ lower UV radiation field  \citep{bro03,abdo10}, thereby allowing us to disentangle
the impact of both factors on the DMG properties. Building on the radial trends discussed in Section
 \ref{sec:2.1}, and the evidence from absorption studies of HCO$^+$ and other molecules in the
direction of the Galactic anti-center showing that large-scale CO surveys are not sensitive to H$_2$
columns associated with the dark neutral medium \citep{liszt2000,liszt23}, we consider their
implications for star formation efficiency across Galactic environments. A large scale survey of OH
with the SKA, covering regions selected from the \lq Outer Galaxy High Resolution Survey'
\citep{colombo21,urquhart25} CO survey of the entire third Galactic plane quadrant ($180 < l < 280$
deg), will be able to pinpoint the importance of CO-dark gas in the outer Galaxy and extend our
understanding of the star formation efficiency trend from the inner to the outer Galaxy.

Turning to the more active inner Galaxy, DMG may also be expected in gas
associated with bar-driven inflow and vertical fountain flows, especially in
the diffuse material feeding or surrounding the CO-bright central molecular
zone (CMZ). The non-circular motions driven by the Galactic bar redistribute
gas inward, while elevated turbulence, shear, and cosmic-ray ionization can
alter the chemical balance of the interstellar medium. In particular, high
cosmic-ray ionization rates can selectively reduce the CO abundance even
where H$_2$ is abundant \citep{2015ApJ...803...37B}, making some fraction of
the inflowing molecular gas difficult to trace with standard CO emission. To understand how the
Galaxy supplies material to its nucleus and regulates star formation there, it is therefore
essential to directly constrain the molecular gas participating in these bar-driven inflows,
including the CO-dark component. Within longitudes $\mid l \mid $ $\leq10^\circ$ and latitudes $\mid
b \mid\leq2^\circ$ \citep[Fig. 2,][]{henshaw2023}, bright \hi\ and CO features
($\mid\textrm{V}_\textrm{LSR}\mid>100$~km\,s$^{-1}$) trace streams of material flowing along the
Galactic bar toward the central molecular zone (CMZ). These inflows play a decisive role in the
growth and evolution of the CMZ, which itself contains the largest reservoir of dense molecular gas
in the Milky Way and is the main driver of nuclear star formation. Although CO emission is clearly
present in these streams \citep{2023ApJ...959...93G,2024ApJ...971L...6S,2024A&A...689A.121V}, a
significant fraction of the molecular component in the central regions is expected to be CO-dark
because the cosmic-ray ionization rate in the CMZ is enhanced by up to 2--3 orders of magnitude
compared to the Solar neighborhood \citep{2025A&A...694A.114R}, leading to the dissociation of CO
and a parallel enhancement of OH through ion–molecule chemistry. As a result, OH absorption becomes
an especially powerful probe of the molecular content in the bar inflows, offering the means to
quantify the fraction of CO-dark H$_2$ in the gas that is actively feeding the CMZ. Indeed, the CMZ
and inflowing streams both show evidence for copious OH absorption in single dish observations
\citep{sawada2004,yan2017,dawson2022}. However, the estimation of CO-dark fractions from these
observations is extremely challenging, since the relative line-of-sight distributions of the
emitting continuum and absorbing molecules are poorly constrained \citep[see][]{dawson2022}. These
difficulties are mitigated for interferometric absorption, since the extended Galactic synchrotron
emission is resolved out, there is far less confusion, and discrete background sources (with
associated distance estimates) can be identified. 

A complementary regime is provided by low and intermediate-latitude fountain flows
\citep[e.g.,][]{2018ApJ...855...33D,2023MNRAS.524.1258N,2023A&A...674L..15V}, where gas is lifted
from the disk by stellar feedback and subsequently cools and returns. These structures contain
moderate metallicity, partially shielded gas where molecules may either survive ejection or reform
during descent. OH absorption toward compact continuum sources at $|b|\leq5^\circ$ allows the
molecular content of vertically circulating gas to be mapped as a function of height above the
plane. This provides a direct test of whether the CO-dark molecular phase is primarily created
during vertical transport or reassembled following cooling, and therefore links feedback-driven
outflows to disk re-accretion and long-term gas cycling. These regions would be covered by the
proposed OH Galactic Plane survey discussed in Section \ref{sec:oh_gp_survey}, and the expected
detection rates in these bright inner Galaxy regions are well represented by the source modeling
presented there. 

Star forming regions in the Galaxy might show a reduced star formation rate even though a
substantial amount of molecular gas is present. Through understanding inner and outer Galaxy
(enhanced and quenched) star formation, we are able to place our local example into its
extragalactic context, enabling a better understanding of the \lq microphysics' driving evolution of
external galaxies. SKA will be essential for estimating the amount of DMG in a variety of
environments and establish its role for star formation and galaxy evolution.

\section{Molecular and atomic lines accessible with SKA}
\label{sec:tracers}

Among the tracers accessible to the SKA, the spectral lines of the simple hydrides OH and CH offer
the most promising probes of molecular gas physics and formation, particularly that of CO-dark gas.
OH and CH are able to trace molecular gas throughout its lifecycle, from the DMG
\citep{wannier1993,barriault2010,allen2012, allen2015, li2018, busch19, Jacob2019, busch2021,
Jacob2021}, through to dense star-forming clouds
\citep[e.g.][]{turner1979,dawson2014,dawson2022,rugel2018,rugel2025,Jacob2024,cappellazzo2025}. A
denser molecular component of the ISM can be characterized in terms of mass and density using
H$_2$CO, providing key advances in understanding of star formation efficiencies in galaxies
\citep{mangum2008,zeiger2010,darling12,GinsburgBally:2015aa,GongOrtiz-Leon:2023rj}. Radio
recombination lines of atoms, primarily H, He, and C, are useful tools to both to pinpoint the
CO-dark gas and to identify the sources of ionization in the warmest components of the ISM. Table
\ref{table:lines} lists the key lines of all of these species available to the SKA.

\begin{table}
\centering
\begin{threeparttable} 
\caption{Frequencies and spectroscopic properties of the key transitions\label{table:lines}} 
\begin{tabular}{l ll c c c c }
\hline \hline 
Species & \multicolumn{2}{c}{Transition} & Frequency  & $E_\mathrm{l}$  & SKA Band \\
& $N, J$ & $F^{\prime} - F^{\prime\prime}$ & [GHz] & [K] & \\
\hline
        \hline
        OH & 1, 3/2& $1^{+}$ -- $2^{-}$ & 1.6122 &  0.00259& Band 2\\
        & & $1^{+}$ -- $1^{-}$ & 1.6654 & 0.00000 & \\
        & & $2^{+}$ -- $2^{-}$& 1.6674 & 0.00259  &\\
        & & $2^{+}$ -- $1^{-}$ &  1.7205 &  0.00000 & \\
        \hline
        CH & 1, 1/2 & $0^{-}$ -- $1^{+}$ & 3.2637 & 0.00072	 & Band 4\\
        &  & $1^{-}$ -- $1^{+}$ & 3.3354 & 0.00072	 & \\
        &  & $1^{-}$ -- $0^{+}$ & 3.3491 & 0.00000   & \\

 & $1, 3/2$ & $2^{-}$ -- $2^{+}$ &  0.7016 & 25.72766 & Band 1\\
       &   & $2^{-}$ -- $1^{+}$ &  0.7039 &  25.72751\\
       &   & $1^{-}$ -- $2^{+}$ &  0.7224 & 25.72766 \\
      &    & $1^{-}$ -- $1^{+}$ &  0.7247 &  25.72751 \\
        \hline
\hi\ & \multicolumn{2}{c}{$^{2}S_{1/2}, F=1$--$0$} &  1.4204 &  0.00000 & Band 1\\
\hline
H RRL$^{a}$ & \multicolumn{2}{c}{265$\alpha$ -- 184$\alpha$} & 0.3513 -- 1.0470 & -- & Band 1\\
H RRL & \multicolumn{2}{c}{190$\alpha$ -- 155$\alpha$} & 0.9512 -- 1.7489 & -- & Band 2\\
H RRL & \multicolumn{2}{c}{112$\alpha$ -- 92$\alpha$} & 4.6188 -- 8.3094 & -- & Band 5a\\
H RRL & \multicolumn{2}{c}{92$\alpha$ -- 75$\alpha$} & 8.3094 -- 15.2815 & -- & Band 5b\\
\hline 
o-H$_2$CO$^{b}$  & \multicolumn{2}{c}{$N_{\rm K_{ a}K_{ c}}$} & \\
 & \multicolumn{2}{c}{$1_{10}$--$1_{11}$} & 4.82966 & 0 & Band 5a\\
  & \multicolumn{2}{c}{$2_{11}$--$2_{12}$} & 14.48848 & 6.75919 & Band 5b\\
\hline
\hline
\end{tabular} 
\begin{tablenotes}
  \item $^{a}$ RRL lines discussed in this Chapter cover Band 2 and 5b; the lines in additional
bands are listed for completeness. $^{b}$ Lower-level energies refer to the ground-state of the
o-H$_2$CO, shifted by $\approx$15 K from the ground-state of p-H$_2$CO. Note: The line frequencies
of the CH transitions were measured in the laboratory by \citet{Truppe2014} with uncertainties of
$\approx$20~Hz, those for the other species were taken from the Cologne Database for Molecular
Spectroscopy \citep[CDMS;][]{Endres2016} and the Jet Propulsion Laboratory  Molecular Database
\citep{jpl}.    
\end{tablenotes}
\end{threeparttable}
\end{table}

\subsection{The OH lines: widespread absorption detections with matched HI absorption}
\label{sec:oh}

The 1.6\,GHz HFS lines of OH are observed throughout the molecular ISM in `quasi-thermal' (i.e.,
non-LTE but not strongly masing) emission and absorption 
\citep{turner1979,dawson2014,dawson2022,rugel2018,rugel2025,Beuther2019}. The low transition
probabilities ($A\sim10^{-11}$\,s$^{-1}$) of these lines combined with the low relative abundance of
OH ($N_{\rm{OH}}/N_{\rm{ H_2}} \sim10^{-7}$) mean that the optical depths of these OH lines are
generally small, and their surface brightness in emission is exceptionally low. Even the brightest
OH `quasi-thermal' emission detections, utilizing sensitive receivers on single-dish radio
telescopes, are typically no more than a few hundred~mK, and may be an order of magnitude fainter in
the CO-dark regime \citep{allen2012,allen2015,busch19,busch2021, busch2024}. Our ability to detect
OH with an interferometer therefore comes primarily from absorption line studies against bright
continuum sources, including point-like extragalactic sources and resolved H{\sc ii} regions  --
both of which are copious at 1.6\,GHz.

As motivated in Section \ref{sec:cnm_to_mgas}, a detailed understanding of the formation of
molecular gas from the cold neutral medium (CNM) requires a statistical census of molecular and
atomic gas properties across a large number of sightlines and a wide range of evolutionary stages.
Comparisons between H{\sc i} and OH absorption can achieve this, allowing us to constrain the
thermal and physical properties of molecular and atomic gas on a component-by-component basis --
including CO-dark diffuse H$_2$. Literature comparisons are limited -- always in sample size, often
in  sensitivity, and also in the degree to which the relationship has been explored.
\citet{hafner2023} measured OH and H{\sc i} absorption toward 38 sightlines using a combination of
data from the Arecibo 305~m telescope and the Australia Telescope Compact Array. They report a CNM
\hi\ column density threshold for molecular gas formation (a result also noted by \citet{li2018},
using largely the same Arecibo sample), but no correlation between the physical properties of their
H{\sc i} and OH components, suggesting a potential decoupling of the atomic and molecular gas
post-transition. \citet{rugel2018} detect OH absorption toward 42 sightlines in the VLA THOR survey,
confirming that OH and CO are generally well correlated, but do not perform a detailed comparison
with H{\sc i}. 
\citet{rugel2025} analyze OH and H{\sc i} absorption together with CO and [C {\sc ii}] emission
toward 3 Galactic H{\sc ii} regions at a sensitivity 5 times better than THOR, discovering only one
completely CO-dark component, but noting moderate molecular gas fractions toward many, and stressing
the importance of deeper observations to widely detect DMG. The GASKAP-OH survey \citep{dawson2024}
on the Australian Square Kilometre Array Pathfinder will reach similar sensitivities, and increase
the sample size via a large Galactic Plane survey, but will still be an order of magnitude shallower
than the SKA survey we propose in Section \ref{sec:survey}. 

The unprecedented sensitivity and resolution of the SKA can provide something that has never before
been achieved: a statistical census of thousands of matched pairs of OH and \hi\ absorption spectra.
Together with complementary CO data from the wealth of existing CO surveys \citep[e.g.,
][]{duarte-cabral21,colombo22}, these can provide the most complete census to date of the cold ISM
pre, post, and during its transition to the molecular phase -- resolved on the sub-parsec spatial
scales on which the key physical processes operate.  
While CO data is an important part of this picture, the combination of matched OH and H{\sc i}
absorption spectra provides a significant advantage: the common background source and high spatial
resolution mean we can be sure that the two tracers are genuinely probing the same volumes of gas (a
problem that plagues comparisons of H{\sc i} and CO, particularly in the confused Galactic Plane).
This means that our ability to assess the association of the two phases will be exquisite.
Furthermore, we will be able to measure the physical state of both the CNM and molecular gas. For
H{\sc i}, the absorption can be combined with off-source emission measurements to return robust
measurements of the physical state of the CNM \citep{mcclure2023}. Although equivalent off-source OH
emission spectra are not a realistic prospect, we have an alternative: the OH line ratios
(particularly in the 1612 and 1720-MHz satellite lines) are sensitively dependent on environmental
conditions, and can be modeled to constrain temperature, density, radiation field and column density
\citep[e.g.][]{guibert1978,vanlangevelde1995,ebisawa19,hafner2020}. We will further outline the
details of a proposed OH Galactic Plane survey in Section \ref{sec:oh_gp_survey}.

\subsection{The ground-, and first excited state lines of CH}
\label{sec:ch}
Like OH, CH serves as a powerful probe of molecular interstellar clouds. Although CH holds the
distinction of being the first molecule detected in space \citep{Swings1937, McKellar1940,
Adams1941} -- via optical absorption near $4300.3$~\AA -- its radio transitions near 3300~MHz,
corresponding to the HFS lines of its ground electronic state, remained underutilized for many years
despite being ubiquitously detected. This was because these lines were almost always observed in
emission, even against bright background sources, and the relative intensities of the three HFS
components often deviated from their expected values under conditions of local thermodynamic
equilibrium (LTE; $I_{3264~{\rm MHz}}$: $I_{3335~{\rm MHz}}$: $I_{3349~{\rm MHz}}$ =$1:2:1$). These
anomalies indicated that the populations of the HFS lines of the CH ground-state $\Lambda$-doublet
levels are inverted. The widespread detection of this population inversion across diverse
environments suggested the presence of a general pumping mechanism that preferentially populates the
upper HFS levels of the ground-state $\Lambda$-doublet, largely independent of local conditions. 

Recently, \citet{Jacob2021} performed detailed non-LTE radiative transfer modeling with the
MOLPOP-CEP code \citep{Asensio2018}, incorporating the effects of preferential pumping from line
overlap. Aided by a combination of HFS–resolved collisional rate coefficients for collisions between
CH and o-/p-H$_2$ \citep{Dagdigian2018}, He \citep{Marinakis2019}, and electrons \citep{Jacob2024},
these authors were able to reproduce the observed intensities and level inversion (see
\citealt{Rygl01.2026.SKA} for more information on CH excitation). The models were further
constrained by column density measurements from CH FIR lines at 2~THz \citep{Jacob2019}. This joint
modeling effort has significantly advanced our understanding of CH excitation and its observational
diagnostics, and it now enables reliable interpretation of CH data even in the absence of
complementary FIR column density estimates-- highlighting that, despite the lack of current SKA
support for S-band observations, pursuing CH remains a particularly fruitful venture given that its
lines are almost always optically thin. Presently the only `large-scale' study of the ground state
lines of CH is being carried out under the framework of the MPIfR-MeerKAT Galactic Plane survey
\citep[MMGPS;][]{Padmanabh2023}. 

In addition to the ground-state lines, the HFS transitions of CH in its first excited state at
700~MHz are particularly important for constraining the pumping scheme of the ground-state maser. To
date, these lines have been detected only in the ISM, toward the W51 giant molecular cloud complex,
first with the Arecibo 305~m telescope \citep{Ziurys1985}, and more recently with the upgraded Giant
Metrewave Radio Telescope (uGMRT) by \citet{Jacob2024}. Jointly modeling these lines with the
ground-state lines has revealed that unambiguous detection of these lines requires targeting regions
with strong 700~MHz continuum emission and moderately high gas densities--sufficiently high to
excite the transition but not exceeding $\sim$10$^{6}$~cm$^{-3}$ (on scales $<$0.5~pc), as is often
the case in high-mass star-forming regions. The principal limitation in detecting this line more
widely has been sensitivity, yet such observations are essential for building a more complete
picture of CH excitation in the ground state. Moreover, the 700~MHz HFS lines offer unique potential
as probes of the line-of-sight magnetic field via the Zeeman effect, owing to CH’s unusually large
Land\'{e} $g$-factor. This diagnostic is especially relevant at gas densities of a few
$\times$10$^{5}$~cm$^{-3}$, comparable to the critical density of the rotational transition
connecting the ground and first excited levels \citep{Bourke01.2026.SKA}. With its unprecedented
sensitivity across Bands 1 and 2 (in the future), the SKA will provide the first real opportunity to
study CH comprehensively in the radio sky, greatly extending its utility as a tracer of CO-dark
molecular gas.

\subsection{o-H$_2$CO: a clean tracer of molecular gas masses and densities}

The lowest two transitions of formaldehyde (o-H$_2$CO at 4.8 and 14.4 GHz) accessible with the
SKA-Mid Band 5 receivers open up a new path to determine gas masses and densities in molecular
clouds. These transitions show anomalous absorption: their collisional excitation drives an
`anti-inversion' which cools the lines below the temperature of the cosmic microwave background
(CMB); the transitions absorb CMB photons and appear in absorption against the CMB
\citep{palmer69,townes69}. Since the CMB is isotropic, the absorption depends only on the total
number of absorbing molecules and is distance-independent; thus, the gas masses can be accurately
determined.

The ratio of the 4.8 and 14.4 GHz lines is a useful tool to determine gas densities, since it is
mostly determined by the volume density of H$_2$ \citep{mangum2008,ginsburg11}. The anomalous
absorption of H$_2$CO lines has been considered to be unaffected by sub-thermal excitation, line
trapping or optical depth effects, which often complicate the interpretation of the emission lines.
However, recent calculations of collisional excitation, including collisions with electrons,
demonstrate that the excitation of the low-frequency ortho-H$_2$CO transitions might depend not only
on H$_2$ density and kinetic temperature, but also on the electron fraction \citep{Gerin2024}. In
particular, the 1$_{10}$--1${_{11}}$ transition at 4.8 GHz is predicted to have an excitation
temperature below the CMB for low to moderate electron fractions ($x(e)<6\cdot10^{-5}$), but to
exceed the CMB at higher electron fractions ($x(e)>10^{-4}$), potentially suppressing or reversing
the anomalous absorption \citep{Gerin2024}. Thus, the interpretation of ortho-H$_2$CO absorption
also depends on the local ionization conditions, at least in the studies which focused on diffuse
and translucent gas.

In Galactic studies, H$_2$CO absorption is typically measured against either the CMB or bright
Galactic continuum sources (e.g., {\hii} regions), probing relatively short and well-defined path
lengths through individual molecular clouds. In contrast, absorption against bright extragalactic
continuum sources such as quasars provides pencil-beam lines of sight that integrate over all
intervening material along the entire path through the Milky Way (and potentially beyond). These
quasar lines of sight therefore probe diffuse and translucent gas components that may be missed in
targeted Galactic observations, but they also introduce line-of-sight averaging over regions with
potentially varying electron fractions, densities, and excitation conditions. As a result, given
these newly computed collisional rate coefficients, the impact of electron excitation on the 4.8 GHz
transition remains to be systematically explored in Galactic environments.

H$_2$CO was successfully detected toward the Galactic Center
\citep{ZylkaGuesten:1992aa,GinsburgBally:2015aa} and several Galactic \hii\ regions
\citep{ginsburg11,GongOrtiz-Leon:2023rj}, as well as star-forming galaxies
\citep{mangum2008,zeiger2010,darling12}. However, the lines are faint and narrow, and therefore not
suitable for large-scale surveys. \citet{GongOrtiz-Leon:2023rj} report the results of the GLOSTAR
survey toward the Cygnus-X star forming region. While widespread absorption is revealed in single
dish Effelsberg maps, the high noise of the interferometric GLOSTAR observations (1.7 K per 0.5
km\,s$^{-1}$ channel) results in absorption on small scales only being detected toward the brightest
continuum sources. Clearly, there is a need for much deeper interferometry to reveal molecular
structures on small scales.   

SKA will be able to detect H$_2$CO in absorption toward a range of environments in the Milky Way,
constraining physical conditions of denser ISM. Combined with OH observations, H$_2$CO lines will
help to establish the role of CO-dark gas for star formation across various density regimes in
molecular clouds, characterized by different metallicities (see Section \ref{sec:starformation}).

In addition, non-anomalous absorption is expected to be detected toward thousands of bright
continuum sources both within the Milky Way and beyond (e.g. radio galaxies and \mbox{H \textsc{ii}}
regions; see
\citealt{Martin-PintadoWilson:1985aa,ArnalGoss:1985ep,KogutSmoot:1989sb,BrunthalerMenten:2021aa}).
Since many \mbox{H \textsc{ii}} regions will have accurately measured trigonometric parallaxes from
associated maser sources \citep{Rygl01.2026.SKA}, it might be possible to conduct 3D tomography of
the molecular ISM by using a network of hundreds of known illuminating sources with
well-characterized heliocentric distances.

\subsection{Radio recombination lines of H, He, and C}
\label{sec:rrl}

RRLs are produced when ions and electrons recombine into an excited state and the electron
transitions toward the ground state. They are the higher principal quantum number corollaries to
H$\alpha$ emission for hydrogen. RRLs can be used to derive plasma properties such as temperature
and density, as well as kinematic distances to the emitting plasma using a Galactic rotation curve
model. Because they are found throughout the radio band, and because their similar properties allow
them to be stacked to increase sensitivity, RRLs are an excellent probe of ionized (for H and He
RRLs) and neutral (for C) gas.

RRL emission does not suffer from extinction and with line-stacking can be quite sensitive
\citep{beuther16}; it is therefore the ideal tool for investigating ionized hydrogen of the WIM and
H\,\textsc{ii}\ regions. Three recently completed single-dish surveys, SIGGMA \citep{liu13, liu19},
GDIGS \citep{anderson21}, and L-band FAST RRL survey \citep{hou22}, show the promise of
fully-sampled RRL maps but lack the sensitivity and angular resolution to fully investigate the
properties and distribution of the WIM.

Because the ionization potential of helium is nearly twice that of hydrogen, the ionic
helium-to-hydrogen ratio, $y^{+}$, indicates the hardness of the radiation field. Within any RRL
observation, the lines of He are shifted $-122$ km s$^{-1}$ from those of H, and so usually fall
within the same bandpass. Observations have shown that $y^{+}$ is similar between the WIM and the
D-WIM \citep{luisi19}, which indicates that the ionizing radiation fields are similar. These
observations have to date been limited to only dense plasmas; sensitive observations are required to
determine $y^{+}$ over the entire Galaxy.

Carbon RRLs are an excellent tracer of the atomic ISM. The ionization potential of carbon is less
than that of hydrogen, so in areas where hydrogen is neutral we still find ionized carbon. C RRLs
are therefore efficient tracers of the properties of photodissociation regions
\citep[PDRs;][]{salas19}, and a useful tool to identify CO-dark gas in low-metallicity regions. Like
He RRLs, C RRLs are also usually in the same bandpass as H RRLs, shifted $-150$ km s$^{-1}$ from
those of H.
 
SKA will enable a large-scale RRL survey of the Galactic midplane at $\sim20\times$ the sensitivity
of GDIGS (see Section~\ref{sec:survey}). This would allow us to trace RRL emission from \hii\
regions into the ISM, test the hypothesis that OB stars provide the requisite photons for
maintaining the WIM, determine the hardness of the radiation field through measurements of $y^+$,
and allow for the determination of PDR properties using C RRLs. It would also allow for a thorough
investigation of the cool WIM seen by \citet{bania24} and its connection to other Galactic features.

\subsection{Note on H\,{\sc i} self-absorption}
\label{sec:hisa}

While details of \hi\ surveys are outside the scope of this chapter, it is worth 
briefly mentioning \hi\ self-absorption (HISA). The SKA will map the \hi\ 21\,cm line at
unprecedented resolution and surface brightness sensitivity, and in particular will provide a new
census of self-absorption from regions where CNM is projected against the warm \hi\ background
emission \citep[e.g.,][]{GibsonTaylor:2000xk,KavarsDickey:2005um}. Observational studies have found
HISA clouds to trace the structure of the CNM \citep[e.g,][]{mcclure2006}, to typically host
molecular gas (as seen via associated tracers such as CO; e.g.  
\citealt{WangBihr:2020cx,SyedBeuther:2023ei}), but also to arise directly from molecule-dominated
regions in the case of the narrowest HISA features \citep[e.g.,][]{LiGoldsmith:2003aa}. Indeed,
recent simulations confirm that HISA may arise not only from the CNM but also from dissociated H$_2$
within self-gravitating molecular clouds, where the latter component is often missed through
observational bias and highly susceptible to the structure of molecular clouds
\citep[][]{SeifriedBeuther:2022un}. We refer the reader to \citet{Miville-Deschenes01.2026.SKA} for a more detailed discussion of \hi\
with the SKA, and here simply highlight that observations of HISA will provide additional
constraints on the structure of the CNM around molecular clouds and on the detailed process of the
\hi\ -to-H$_2$ transition.

\section{Proposed observations}\label{sec:survey}

\subsection{Galactic plane survey in OH}
\label{sec:oh_gp_survey}

As discussed in Section \ref{sec:oh}, an SKA Galactic Plane survey in OH absorption -- when paired
with an equivalent H{\sc i} survey -- would provide a statistical census of thousands of matched
pairs of OH and H{\sc i} absorption
spectra. Such a dataset would allow us to statistically track how the presence and properties of the
molecular gas vary with CNM properties and with $R_{\rm Gal}$ and local Galactic environment. We
would be able to pin down the CNM conditions required for molecule formation, including
distinguishing DMG from CO-bright H$_2$. We would be able to assess correlations between \hi\ and
H$_2$ physical properties, measuring the imprint of the CNM upon the molecular gas formed within it.
We would be able to assess correlations both spatially (across resolved background sources) and
spectrally (e.g., profile shapes, linewidths) to constrain the degree of mixing of the molecular and
atomic phase, and compare our results with the predictions of H$_2$ formation models
\citep[e.g.][]{goldsmith2007, sternberg14, seifried17, bialy2017} to comment on key properties such
as the turbulent properties and degree of clumping of the host CNM \citep{park2023}. These analyses
coming together, in a statistical way across varied conditions that exist in the plane of the Milky
Way, could constitute a generational change in our understanding of the transitions between the key
phases and gas populations in our Galaxy.
 \begin{figure}
     \centering
    \includegraphics[width=0.6\linewidth]{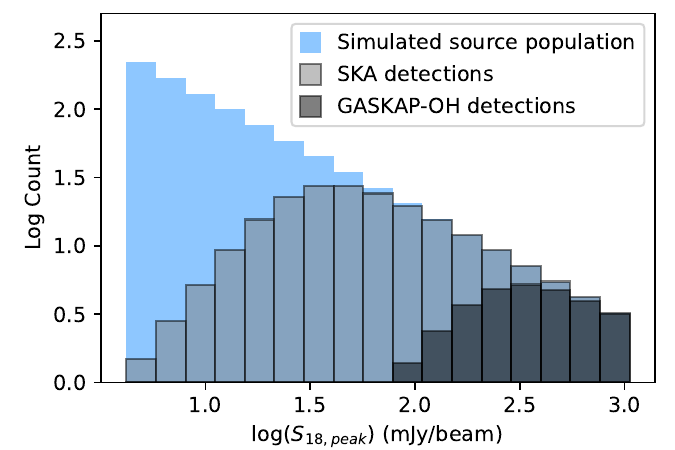}
     \caption{Simulated detection rate for OH absorption over a 600 square-degree SKA survey of the
Galactic Plane. The simulated data is generated as described in the text. This treatment considers
only the brightest position toward a simulated source, and is in that sense a lower limit. For
extended bright sources, multiple/resolved detections are likely; for extended weak sources,
integrating spatially to improve sensitivity could yield detections not modeled here. Conversely,
the source population is modeled from a relatively bright region of the Galactic Plane (likely
overestimating compared to a larger survey) but extending to $|\,b\,|\lesssim2.1^{\circ}$ -- higher
than an SKA survey likely would. While the exact numbers here are therefore subject to some
uncertainty, the order of magnitude detection rate and improvement over GASKAP-OH (the closest
competitor) are reasonable.}
     \label{fig:oh_detections}
 \end{figure}
 
The sensitivity of OH absorption measurements is driven by the flux density distribution of the
background continuum source population. In the Galactic Plane this consists of both compact
extragalactic sources and resolved H{\sc ii} regions and supernova remnants -- the latter of which
often dominate the bright source population, particularly in the inner Galaxy. 
here that while off-Plane and outer-Galaxy sightlines offer advantages in terms of spectral
simplicity,  the density of sufficiently bright background sources is far lower.) 
The measured flux density distribution of resolved sources is strongly affected by $uv$ coverage and
imaging parameters and is therefore not straightforward to predict. 
The expected optical depth distribution of OH in the Galactic Plane is similarly not well
constrained, but likely ranges from $\lesssim0.01$ \citep[as seen in very diffuse, off-Plane
sightlines;][]{li2018, NguyenDawson:2018xt, hafner2023} to $\tau\sim1.0$ as seen toward some
star-forming gas associated with \hii\ regions
\citep[e.g.,][]{rugel2018,KoleyRoy:2021zw,cappellazzo2025}. In order to constrain both of these
distributions and make meaningful predictions about detection rates, we turn to Australian SKA
Pathfinder (ASKAP) data from the GASKAP-OH survey \citep{dawson2024}. With baselines ranging from
22m to 6km and fairly dense \textit{uv} coverage, particularly on shorter baselines, the appearance
of the complex in-Plane continuum in ASKAP data is likely to be comparable to that of the AA4 array
(though the imaging fidelity with AA4 should be superior). \citet{dawson2024} report the OH
detection rate as a function of optical depth sensitivity ($\sigma_{e^{-\tau(v)}}$) for a single
$\approx20$ square-degree GASKAP-OH field centered on $l\approx340^{\circ}$ and extending between
$|\,b\,|\lesssim2.1^{\circ}$. We take the continuum source peak flux density distribution ($S_{\rm
peak}$) from the same data\footnote{Extracted from contiguous `islands' of emission identified by
the \textit{Selavy} sourcefinder  \citep{whiting2012}}, which are well described by a simple power
law, and extrapolate down to a flux density limit of 5\,mJy. We assume an on-source time of 600s and
adopt reasonable imaging parameters to obtain a spectral sensitivity at 1.6 GHz of
$\sigma_v\sim1$\,mJy for a velocity resolution of 0.75\,km\,s$^{-1}$. With optical depth sensitivity
defined as $\sigma_{e^{-\tau(v)}}=\sigma_v/S_{\rm peak}$, we thus can obtain a simulated source
population and detection statistics for a hypothetical SKA survey.

Figure \ref{fig:oh_detections} shows the results of this simulation, together with a description of
the caveats. In the inner Galaxy where the on-sky source density of H{\sc ii} regions and SNRs is
high, we expect to detect OH absorption toward at least $\sim10$ sources/deg$^2$ (many with multiple
spectral components), 
and to produce \textit{resolved maps} of OH optical depth toward the brighter continuum complexes,
enabling us to track small-scale variations in gas properties at a resolution of $\sim4\arcsec$
($\sim0.05$ pc at a representative distance of 3\,kpc). In the outer Galaxy where the source
distribution will trend toward the (generally weaker) extragalactic population, and the OH optical
depth may be lower, the detection rate is expected to be correspondingly smaller. Alongside an
unbiased survey of the inner Galaxy, deeper integrations toward targeted outer Galaxy fields would
therefore be a more efficient use of telescope time.

Although the 3~GHz (Band 4) receiver remains unfunded at present, the exceptional sensitivity of the
SKA offers a unique opportunity to probe the 700~MHz CH lines on a large scale for the first time.
The excitation of these CH transitions is primarily dominated by collisions, and as a result they
are expected to appear in absorption against the bright background envelopes surrounding the
targeted star forming regions \citep[see Sect.~5.3 of][]{Jacob2024}. Toward W51~E (the only known
source toward which this line has been unambiguously detected), \citet{Jacob2024} detect absorption
that is only a few \% of the continuum. Therefore, a primary reason for its non-detection thus far
has been due to sensitivity issues arising from the weaker background continuum emission at 700~MHz.
SKA will allow us to probe down to 5~mJy, sufficient to detect CH absorption at a minimum of a
3~$\sigma$ level (based on previous observations).

\subsection{\hii\ regions}
There are 36 H$\alpha$ RRLs available in Band 2 and 17 in Band 5b. Using the SKA sensitivity
calculator with an integration time of 600 s 
and assuming a robust weighting of 1 gives a sensitivity of $\sim$0.7\,mJy\,beam$^{-1}$ per
3.5\,km\,s$^{-1}$ channel in Band 2 and $\sim$1\,mJy\,beam$^{-1}$ per 4.5\,km\,s$^{-1}$ channel in
Band 5b, with beam sizes of $\sim$1 arcsec and $\sim$0.1 arcsec, respectively. 

\begin{figure}[!ht]
\centering 
\includegraphics[width=0.90\textwidth]{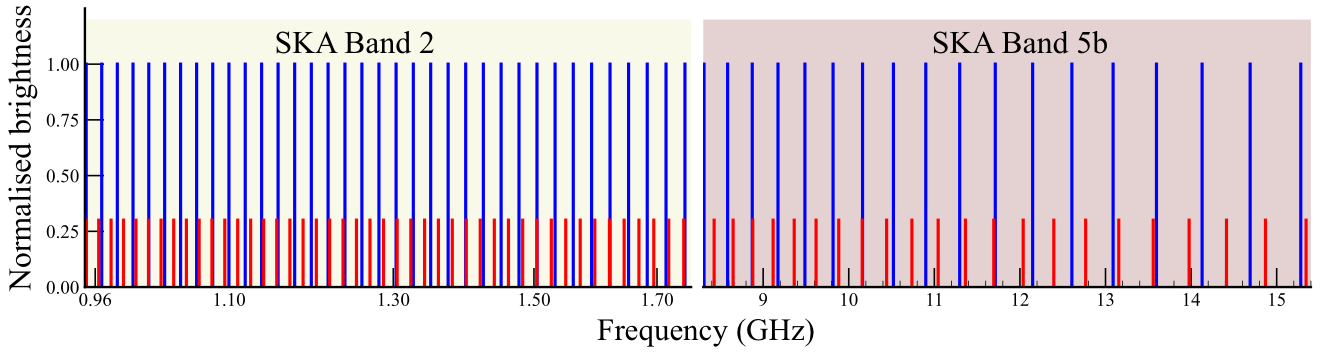}
\caption{Schematic showing the RRLs within the SKA Bands 2 and 5b. The blue and red lines show the
frequencies of the H$\alpha$ and H$\beta$ lines respectively. There are 36 H$\alpha$ RRLs in Band 2
(beige shaded region) and 17 H$\alpha$ RRLs in band 5b (pink shaded region).}
\label{fig:rrl_schematic}
\end{figure}

The sensitivities in Band 2 can be improved by stacking RRLs (Figure \ref{fig:rrl_schematic}). This
is also possible in Band 5b, but the number of available lines within each 2.5 GHz window is much
smaller, and the process is more complicated because the line intensity depends on frequency.
However, the sensitivity in Band 5b can be straightforwardly enhanced by smoothing the data to a
1-arcsec beam.

\begin{figure}
\centering 
\includegraphics[height=0.33\textwidth]{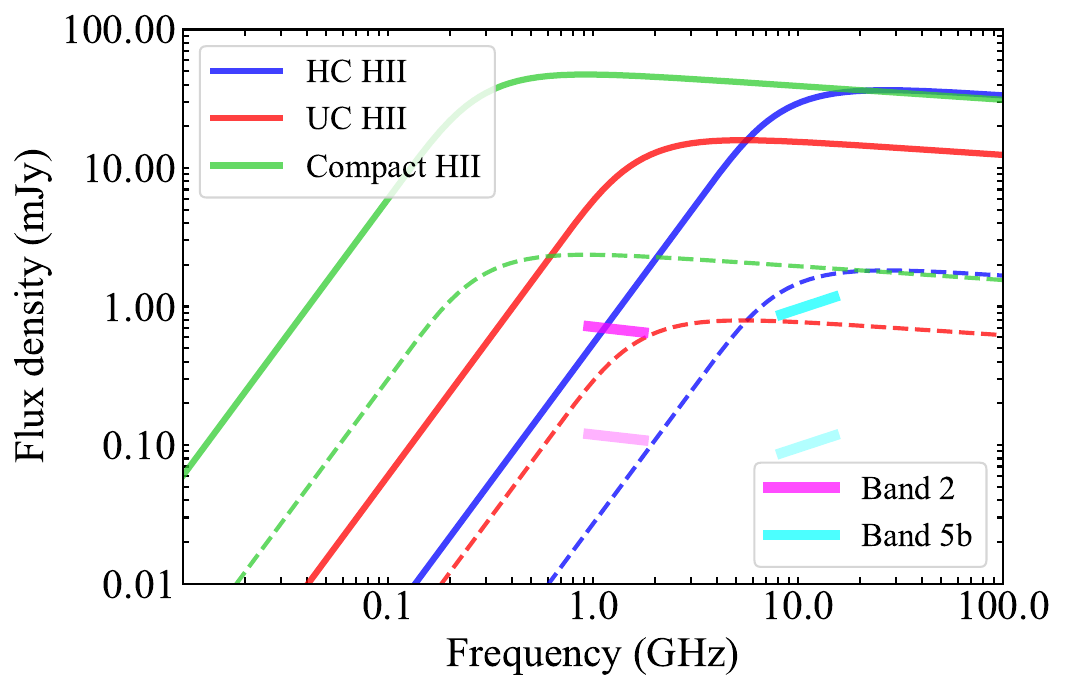}
\includegraphics[height=0.33\textwidth]{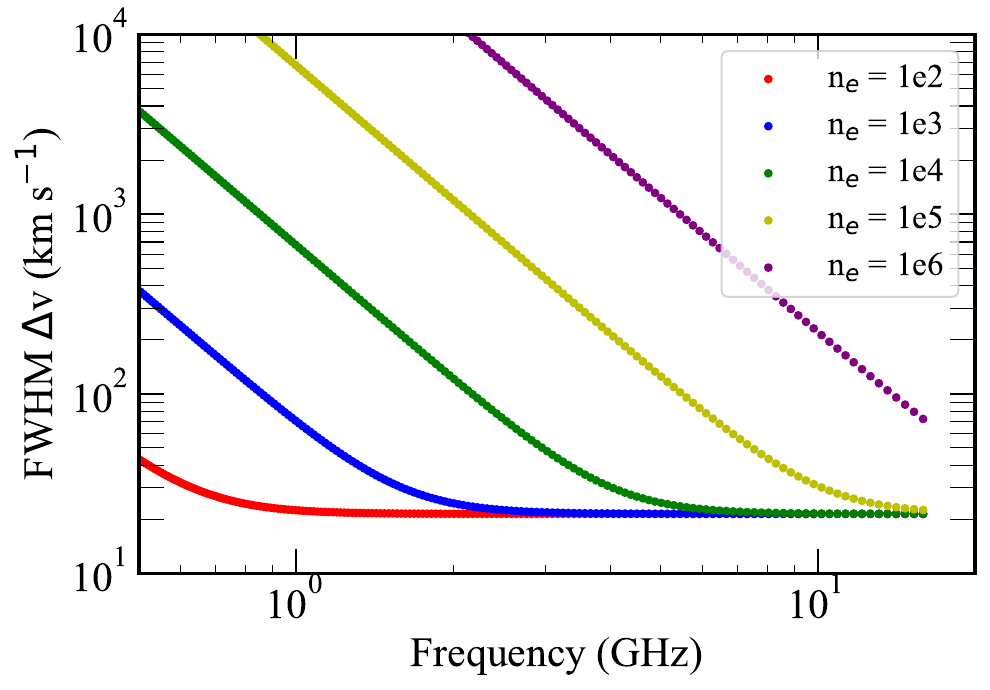}
\caption{Left Panel: Predicted flux densities for the three most compact \hii\ region stages. The
solid and dashed lines show the continuum and RRL fluxes respectively. The bright magenta and cyan
bars show the sensitivity to RRLs estimated from the SKA sensitivity calculator assuming a 600
second integration, robust weighting of 1 and channel width of $\sim$3.5 and
$4.5\,\mathrm{km\,s^{-1}}$. The pale magenta and cyan bars show the improvement to the sensitivity
from stacking the RRLs in Band 2 and smoothing to 1-arcsec resolution in Band 5b. Right panel: Shows
the impact of pressure broadening to the expected RRL line-width as a function of electron density
and frequency.}
\label{fig:hii_fluxes}
\end{figure}

In Figure \ref{fig:hii_fluxes}, we show the expected continuum and RRL fluxes for the three most
compact stages of \hii\ regions located at a distance of 20 kpc, assuming a line-to-continuum flux
ratio of 20. We also show the RRL sensitivities (brighter magenta and cyan bars) as well as the
improved sensitivities achievable by stacking and/or smoothing the data. These results demonstrate
that the SKA will be capable of detecting RRLs from all compact \hii\ region stages; however,
detecting the two most compact stages will be challenging in Band 2.

A further complication arises from pressure broadening, which is significant for compact \hii\
regions with high electron densities, particularly at low frequencies. This effect is especially
severe in Band 2, leading to line widths of hundreds of km\,s$^{-1}$ for moderate electron densities
($\sim 10^4$\,cm$^{-3}$), and may limit the usefulness of Band 2 to the more extended compact \hii\
regions. In contrast, pressure broadening is much less severe in Band 5b, which is better suited for
probing the most compact \hii\ regions. However, Band 5b is less sensitive to more extended \hii\
regions, which are more easily filtered out.

The two bands are therefore complementary: Band 2 is more sensitive to larger \hii\ regions with
lower electron densities, while Band 5b is more sensitive to smaller, denser \hii\ regions.
Combining observations from both bands will be essential for achieving a complete census of the
compact \hii\ region population. We note there are a further 21 H$\alpha$ RRLs in Band 5a (i.e.,
4.6-8.5\,GHz) that can enhance the analysis and provide additional constraints on the evolutionary
modeling  but the focus of this chapter is on Band 2 and 5b.  

\subsection{Physical conditions of dense ISM with H$_2$CO}

Formaldehyde absorption at 4.8 and 14.4\,GHz has been mapped with single-dish surveys for several
regions in the Milky Way down to angular scales of $\geq1'$
\citep[e.g.,][]{ZylkaGuesten:1992aa,GinsburgBally:2015aa,GongOrtiz-Leon:2023rj}, but wide area
surveys at much higher angular resolution are more challenging. For example, to detect formaldehyde
absorption lines from Galactic clouds requires a sensitivity of $\sim$0.1-0.2 K, which takes at
least a full 12-hour track with the VLA, yielding a synthesized beam of ${21'' \times 18''}$ at
$\sim$4.8~GHz \citep{evans87}. For comparison, the wide area GLOSTAR JVLA survey reaches an rms
sensitivity of 1.7 K per 0.5 km\,s$^{-1}$ channel for the 4.8 GHz formaldehyde line and so can only
reveal the brightest  components \citep{GongOrtiz-Leon:2023rj}.

The SKA-Mid will fundamentally revolutionize this type of investigation across the Galactic plane,
combining sensitive and high angular resolution observations with a wide Field of View
($\sim$12.5$'$ and $\sim$6.7$'$ in Bands $5a$ and $5b$, respectively;
\citealt{braun2019anticipatedperformancesquarekilometre}). For instance, exploiting Band $5a$
receivers ($\nu_0 =$~4.8~GHz, see Tab.~\ref{table:lines}; $\Delta\nu = 6.72 \text{ kHz}$), an rms
noise level of approximately 0.2~K 
can be achieved in  $\sim20$ hours using the $\text{AA}4$ configuration, yielding a synthesized beam
of $\sim 7''$ (Briggs; $\text{robust}=0$), considerably smaller than the $\sim 20''$ beam  of VLA
observations at these frequencies \citep{evans87}. This sensitivity will allow us to spectrally
resolve the $\text{H}_2\text{CO}$ $1_{10}$--$1_{11}$ lines, which have a typical $\text{FWHM}$ of
$\sim 1 \text{ km s}^{-1}$, as reported by \citet{evans87}. In a comparable amount of time, an
$\text{H}_2\text{CO}$ $2_{11}$--$2_{12}$ detection will be feasible using the $\text{SKA-Mid Band }
5b$ receiver ($\nu_0 =$~14.5~GHz, see Tab.~\ref{table:lines}; $\Delta\nu = 13.44 \text{ kHz}$),
achieving an rms of $\sim0.1$~K with a synthesized beam of $\sim 3.5''$ (Briggs; $\text{robust}=1$).
This capability makes it feasible to survey the Milky Way’s molecular clouds at significantly better
angular resolution than existing single-dish $\text{CO}$ surveys, and without the excitation or
optical depth issues that typically affect those $\text{CO}$ observations.

\section{Summary and outlook}
\label{sec:summary}

The SKA will open a new discovery space in Galactic ISM studies by enabling wide-area, high-fidelity
spectroscopic surveys that combine deep continuum mapping with simultaneous spectral line
observations at high velocity resolution. By leveraging these capabilities, the proposed OH
absorption line survey of the Galactic Plane, combined with matched \hi\ data, will deliver the
first large-scale statistical census of OH–\hi\ absorption pairs. This dataset will quantify how
molecular gas emerges from the CNM, including identifying the physical conditions that regulate
H$_2$ formation, distinguishing CO-bright and CO-dark molecular components, and directly measuring
how the CNM structure and dynamics imprint on newly formed molecular gas. A future pathway for 3 GHz
(Band 4) CH absorption line measurements will provide a complementary and highly sensitive tracer of
diffuse molecular gas when this SKA receiver becomes available.

At the same time, targeted RRL observations across Bands 2 and 5b will provide a comprehensive view
of the compact \hii\ region population, tracing the evolution of ionized gas from the densest
hyper-compact objects through to extended compact regions. The complementarity of the two bands --
with Band 2 optimized for more extended, lower-density regions, and Band 5b sensitive to the
youngest, densest regions less affected by pressure broadening -- will be critical to constructing a
complete evolutionary census. When combined with continuum survey products, even shallow
per-pointing integrations can yield powerful RRL constraints through line stacking and beam-matched
smoothing, demonstrating the efficiency of commensal survey strategies.
Together, these experiments will deliver:
(i) the largest absorption-based inventory of molecular gas tied directly to CNM properties;
(ii) spatially resolved maps of OH opacity across key continuum complexes, probing sub-pc-scale ISM
structure;
(iii) empirical constraints on H$_2$ formation models under realistic Galactic conditions; and
(iv) a unified, multi-band RRL census of compact \hii\ regions across the Milky Way.

More broadly, these observations showcase the unique impact of SKA’s multi-line, wide-band survey
mode. Building on the legacy of VLA pathfinder programs such as THOR, GLOSTAR, and the L-band Local
Group Survey, an SKA survey designed for commensal continuum and spectral line science will not only
maximize discovery potential, but will also directly inform optimal correlator configurations,
survey cadences, and stacking strategies for next-generation ISM studies.

By simultaneously charting the molecular, atomic, and ionized phases of the Milky Way with
unprecedented sensitivity and statistical power, the SKA will enable a step change in our
understanding of phase transitions in the Galactic ecosystem -- from how molecules first assemble in
the diffuse ISM to how massive stars reshape their natal environments. This integrated view has the
potential to redefine Galactic ISM studies for decades to come, while providing essential benchmarks
for star-forming galaxy physics across cosmic time.

\section*{Acknowledgments}
\footnotesize{A.~K. acknowledges support from the Polish National Science Center SONATA BIS grant
No. 2024/54/E/ST9/00314. A.~M.~J. and V.~V.~S. would like to acknowledge the support of the Max
Planck Society. D.~C., A.~K., A.~M.~J., and V.~V.~S. gratefully acknowledge the Collaborative
Research Center 1601 (SFB 1601 sub-project B3, A1, and B2) funded by the Deutsche
Forschungsgemeinschaft (DFG, German Research Foundation) – 500700252. M.P.B. is supported by a
Jansky Fellowship provided by the National Radio Astronomy Observatory. The National Radio Astronomy
Observatory and Green Bank Observatory are facilities of the U.S. National Science Foundation
operated under cooperative agreement by Associated Universities, Inc. G.~S. acknowledges financial
support under the National Recovery and Resilience Plan (NRRP), Mission 4, Component 2, Investment
1.1, Call for tender No. 104 published on 2.2.2022 by the Italian Ministry of University and
Research (MUR), funded by the European Union – NextGenerationEU-Project Title 2022JC2Y93 Chemical
Origins: linking the fossil composition of the Solar System with the chemistry of protoplanetary
disks – CUP J53D23001600006 – Grant Assignment Decree No. 962 adopted on 30.06.2023 by the Italian
Ministry of Ministry of University and Research (MUR); the project ASI-Astrobiologia 2023 MIGLIORA
(“Modeling Chemical Complexity”, F83C23000800005); the INAF-GO 2024 fundings ICES, the INAF-GO 2023
fundings PROTOSKA (“Exploiting ALMA data to study planet forming disks: preparing the advent of
SKA”, C13C23000770005) and the INAF Minigrant 2023 TRIESTE (“TRacing the chemIcal hEritage of our
originS: from proTostars to planEts”; PI: G. Sabatini). V.V.S acknowledges the support of the
Department of Atomic Energy, Government of India, under Project Identification No. RTI 4012.}

\clearpage

\bibliographystyle{abbrvnat-maxbibnames4}
\bibliography{chapter}

\end{document}